\documentclass[]{pasj02}
\usepackage[switch,mathlines]{lineno} 

\jyear{2026}
\Received{}
\Accepted{}

\newcommand{\nustar}{\textit{NuSTAR}}
\newcommand{\xrism}{\textit{XRISM}}
\newcommand{\xtend}{\textit{Xtend}}
\newcommand{\resolve}{\textit{Resolve}}
\newcommand{\xmmnewton}{\textit{XMM-Newton}}
\newcommand{\rgs}{\textit{RGS}}
\newcommand{\nicer}{\textit{NICER}}
\newcommand{\swift}{\textit{Swift}}
\newcommand{\seimei}{\textit{Seimei}}
\newcommand{\suzaku}{\textit{Suzaku}}
\newcommand{\xis}{\textit{XIS}}

\newcommand{\revisone}[1]{#1}
\newcommand{\revistwo}[1]{#1}
\newcommand{\revistwomaj}[1]{#1}
\newcommand{\revisthree}[1]{#1}

\newcommand{\revis}[1]{#1}

\usepackage[authoryear]{natbib}
\usepackage{siunitx}
\DeclareSIUnit\erg{erg}
\usepackage{physics}
\usepackage{multirow}
\usepackage{ulem}
\usepackage{rotating}
\AtBeginDocument{\RenewCommandCopy\qty\SI}

\begin{document}

\title{The Geometry of Ultra-Fast Outflows Probed by Soft X-ray Variability in PDS~456}

\author{
 Riki \textsc{Sato},\altaffilmark{1}\altemailmark\orcid{0009-0002-6943-8169} \email{riki.sato@phys.s.u-tokyo.ac.jp}
 Kouichi \textsc{Hagino},\altaffilmark{1}\orcid{0000-0003-4235-5304}
 Toshiya \textsc{Iwata},\altaffilmark{1,2}\orcid{0000-0002-2304-4773}
 Ehud \textsc{Behar},\altaffilmark{3}\orcid{0000-0001-9735-4873}
 Valentina \textsc{Braito},\altaffilmark{4,5,6}
 Chris \textsc{Done},\altaffilmark{7}
 Misaki \textsc{Mizumoto},\altaffilmark{8}
 James N. \textsc{Reeves},\altaffilmark{5,4}
 Rozenn \textsc{Boissay-Malaquin},\altaffilmark{9,10,11}
 Pierpaolo \textsc{Cond\'o},\altaffilmark{12}
 Luigi C. \textsc{Gallo},\altaffilmark{13}
 Adam G. \textsc{Gonzalez},\altaffilmark{13}
 Alfredo \textsc{Luminari},\altaffilmark{14,15}
 Aiko \textsc{Miyamoto},\altaffilmark{16}
 Ryuki \textsc{Mizukawa},\altaffilmark{17}
 Hirokazu \textsc{Odaka},\altaffilmark{16}
 Atsushi \textsc{Tanimoto},\altaffilmark{18}
 Makoto \textsc{Tashiro},\altaffilmark{17}
 Francesco \textsc{Tombesi},\altaffilmark{12,15,19}
 Yerong \textsc{Xu},\altaffilmark{13,20,21}
 Satoshi \textsc{Yamada},\altaffilmark{22,23,24}
 Tahir \textsc{Yaqoob},\altaffilmark{9,10,11}
 Aya \textsc{Bamba},\altaffilmark{1,25,26}\orcid{0000-0003-0890-4920}
}
\altaffiltext{1}{Department of Physics, The University of Tokyo, 7-3-1 Hongo, Bunkyo-ku, Tokyo 113-0033, Japan}
\altaffiltext{2}{RIKEN Pioneering Research Institute, 2-1 Hirosawa, Wako, Saitama 351-0198, Japan}
\altaffiltext{3}{Department of Physics, Technion, Technion City, Haifa 3200003, Israel}
\altaffiltext{4}{INAF, Osservatorio Astronomico di Brera, Via Bianchi 46, I-23807 Merate (LC), Italy}
\altaffiltext{5}{Department of Physics, Institute for Astrophysics and Computational Sciences, The Catholic University of America, 620 Michigan Ave., N.E., Washington, DC 20064, USA}
\altaffiltext{6}{Dipartimento di Fisica, Universit\`a di Trento, Via Sommarive 14, Trento, Italy}
\altaffiltext{7}{Centre for Extragalactic Astronomy, Department of Physics, University of Durham, South Road, Durham DH1 3LE, UK}
\altaffiltext{8}{Science Research Education Unit, University of Teacher Education Fukuoka, 1-1 Akamabunko-machi, Munakata, Fukuoka 811-4192, Japan}
\altaffiltext{9}{Center for Space Science and Technology, University of Maryland, Baltimore County (UMBC), 1000 Hilltop Circle, Baltimore, MD 21250, USA}
\altaffiltext{10}{NASA / Goddard Space Flight Center, Greenbelt, MD 20771, USA}
\altaffiltext{11}{Center for Research and Exploration in Space Science and Technology, NASA / GSFC (CRESST II), Greenbelt, MD 20771, USA}
\altaffiltext{12}{Physics Department, Tor Vergata University of Rome, Via della Ricerca Scientifica 1, I-00133 Rome, Italy}
\altaffiltext{13}{Department of Astronomy \& Physics, Saint Mary's University, 923 Robie St, Halifax, Nova Scotia B3H 3C3, Canada}
\altaffiltext{14}{INAF, Istituto di Astrofisica e Planetologia Spaziali, Via del Fosso del Cavaliere 100, I-00133 Rome, Italy}
\altaffiltext{15}{INAF, Osservatorio Astronomico di Roma, Via Frascati 33, I-00078 Monte Porzio Catone, Italy}
\altaffiltext{16}{Department of Earth and Space Science, Graduate School of Science, \revistwo{The University of Osaka}, 1-1 Machikaneyama, Toyonaka, Osaka 560-0043, Japan}
\altaffiltext{17}{Department of Physics, Saitama University, 255 Shimo-Okubo, Sakura, Saitama 338-8570, Japan}
\altaffiltext{18}{\revistwo{Department of Informatics, Osaka Gakuin University, 2-36-1 Kishibeminami, Suita-City, Osaka 564-8511, Japan}}
\altaffiltext{19}{\revistwo{INFN, Rome Tor Vergata, Via della Ricerca Scientifica 1, I-00133 Rome, Italy}}
\altaffiltext{20}{\revistwo{Institute of Space Sciences (ICE, CSIC), Campus UAB, Carrer de Magrans, 08193 Barcelona, Spain}}
\altaffiltext{21}{\revistwo{Institut d'Estudis Espacials de Catalunya (IEEC), Edifici RDIT, Campus UPC, 08860 Barcelona, Spain}}
\altaffiltext{22}{Frontier Research Institute for Interdisciplinary Sciences, Tohoku University, Aramaki, Aoba-ku, Sendai, Miyagi 980-8578, Japan}
\altaffiltext{23}{\revistwo{Department of Astronomy, University of Geneva, ch. d'Ecogia 16, 1290, Versoix, Switzerland}}
\altaffiltext{24}{\revistwo{Astronomical Institute, Tohoku University, 6-3 Aramakiazaaoba, Aoba-ku, Sendai, Miyagi 980-8578, Japan}}
\altaffiltext{25}{Research Center for the Early Universe, School of Science, The University of Tokyo, 7-3-1 Hongo, Bunkyo-ku, Tokyo 113-0033, Japan}
\altaffiltext{26}{Trans-Scale Quantum Science Institute, The University of Tokyo, 7-3-1 Hongo, Bunkyo-ku, Tokyo 113-0033, Japan}




\maketitle

\begin{abstract}
Constraining the location and geometry of ultra-fast outflows (UFOs) is essential for identifying where they are launched and how they are accelerated. We investigate the soft X-ray variability of the luminous quasar PDS~456 using simultaneous March 2024 observations with \xrism/\xtend\ and \nustar. A model-independent comparison between the flare and quiescent phases reveals spectral variability around 1~keV in the rest frame, while the hard X-ray spectral shape remains nearly unchanged. Broadband spectral fitting shows that the soft X-ray structure is well described by a partial-covering low-ionization UFO with $\log (\xi/(\unit{\erg\cm\per\s})) \simeq 3.1$ and $v_\mathrm{out}\simeq 0.30c$. Time-sliced spectral analysis further reveals significant variability in the covering fraction of this absorber. Interpreting this variability as transverse motion across the X-ray source, we constrain the crossing velocity to be $v_\mathrm{cross}\lesssim \num{5e-3}c$ and derive a lower limit on the absorber distance of $r \gtrsim \num{4e3}\,R_\mathrm{g}$. This location is substantially farther out than the high-ionization UFO previously inferred at $\sim 200$--$600\,R_\mathrm{g}$, \revisone{while} the two phases have comparable outflow velocities. \revistwo{The resulting velocity--distance structure disfavors a self-similar magnetocentrifugal wind and instead suggests either radiation-pressure acceleration following a Castor-Abbott-Klein-like velocity law or compact magnetic acceleration through magnetic reconnection.} These results demonstrate that soft X-ray partial-covering variability can provide a geometrical probe of UFOs and directly connect spectral variability to wind acceleration.
\end{abstract}


\section{Introduction}
Observational correlations between supermassive black holes (SMBHs) and host-galaxy properties, such as the $M$--$\sigma$ and SMBH--bulge relations, indicate that SMBH growth is closely linked to galaxy evolution \citep {Magorrian1998, Ferrarese2000, Kormendy2013}.
This connection is commonly interpreted in terms of AGN feedback \citep {Silk1998,King2003,Fabian2012a}, in which energy and momentum released by accreting SMBHs affect the surrounding gas and can regulate star formation.
Among the possible feedback channels, ultra-fast outflows (UFOs) are particularly important \revis{\citep{Tombesi2012, KingPounds_Feedback2015}}.
\revistwo{UFOs are observed in X-ray spectra as blueshifted absorption features from highly ionized gas in both the Fe--K and soft X-ray bands, with velocities of $\sim 0.03$--$0.3c$ \citep[e.g.,][]{Pounds2003,Tombesi2010,Gofford2013,Serafinelli_2019,SUBWAYS2023}.} However, the physical mechanism by which UFOs are launched from the accretion disk and accelerated to mildly relativistic velocities remains uncertain. \par
Two representative classes of acceleration mechanisms have been widely discussed: radiation-driven winds and magnetohydrodynamic (MHD) winds. \revisone{Radiation pressure has been \revistwo{proposed} as a mechanism for launching and accelerating UFOs through both continuum and line opacity.
In accretion flows above the Eddington limit, continuum radiation pressure due to electron scattering can contribute significantly to driving outflows \citep[e.g.,][]{King2003,Takeuchi2013}. In sub-Eddington flows, where electron scattering alone is insufficient, bound-bound line opacity can instead enhance the radiative force, as in the classical line-driving framework of \citet{CAK_1975}.}
\revistwo{In line-driven UFO scenarios, however, the gas must avoid being overionized for UV line driving to operate efficiently. X-ray shielding, which depends on the wind geometry, may help explain highly ionized absorption in such winds \citep{Hagino2015, Nomura2016,Mizumoto2021}.}
In MHD-driven models, large-scale magnetic fields extract angular momentum from the accretion disk and accelerate gas along magnetic field lines, as in magnetocentrifugal or self-similar MHD disk-wind solutions \citep [e.g.,][]{Blandford1982,Contopoulos1994,Fukumura2010}. More general magnetic-acceleration scenarios, such as compact magnetic-energy dissipation through magnetic reconnection, have also been proposed from recent observations \citep {Gu2025a}.
Although \revisone{both radiation-driven and MHD-driven scenarios} can account for some observed properties of UFOs \revistwo{\citep[e.g.,][]{Mizumoto2021,Fukumura2015}}, the dominant launching mechanism remains uncertain. Because \revisone{the two mechanisms} predict different velocity--radius relations and wind geometries, observational constraints on the location and geometry of UFOs \revistwo{enable us to distinguish} between them. \par
\revisone{\revistwo{PDS~456 is a nearby luminous quasar at $z=0.184$ \citep{Torres_PDS456_red_1997}, and} is well suited for studying UFO geometry. With a bolometric luminosity of $\sim \qty{e47}{\erg\per\s}$ and a black hole mass of $\sim\num{5e8}~M_\odot$ \citep{GravityCollab_2023}, PDS~456 is likely accreting at or above the Eddington limit. }
\revisone{It} is well known for hosting powerful UFOs, which have been extensively studied through ionized Fe-K absorption in the 8--10~keV rest-frame band. \revis{These studies show that the velocity of the high-ionization UFO in PDS~456 is typically $\sim 0.25$--$0.3c$ and that its ionization parameter is high, with $\log (\xi/(\unit{\erg\cm\per\s})) \sim 5$--$6$ \citep[e.g.,][]{Reeves2003,Nardini2015,Matzeu2016}.} Here, the ionization parameter is defined as $\xi \equiv L_{\rm ion}/(n r^2)$, where $L_{\rm ion}$ is the luminosity in the ionizing band (1--1000~Ryd), $n$ is the \revistwo{electron} number density, and $r$ is the distance from the ionizing source to the gas. In addition, previous observations have suggested the presence of a low-ionization UFO component in the soft X-ray band below $\sim$3~keV, with a velocity of $\sim 0.2$--$0.3c$ and $\log (\xi/(\unit{\erg\cm\per\s})) \sim 3$--$4$ \citep [e.g.,][]{gofford_2014_lowUFO,reeves_2016_low_UFO}.
\revistwo{However, the location of this low-ionization UFO remains poorly understood. Several geometries are possible, including the X-ray-shielded interiors of high-ionization UFO clumps, a more distant phase along the same streamline, or an independent outflow launched from a different disk radius. A distance constraint on the low-ionization UFO would allow comparison with the high-ionization UFO on the velocity--distance plane. Their velocity--distance relation can provide a direct clue to the acceleration mechanism.} \par
\revistwo{In March 2024, PDS~456 was observed simultaneously with \xrism, \nustar, \xmmnewton, \nicer, \swift, and \seimei, together covering the optical, ultraviolet, and X-ray bands. \citet{Firstpaper2025} studied the high-ionization UFO in the Fe-K band using \xrism/\resolve. Owing to the high energy resolution of \xrism/\resolve, they resolved the Fe-K absorption into multiple velocity components and interpreted this structure as evidence for a clumpy wind. This clumpy-wind picture allowed them to estimate the characteristic clump properties and constrain the location of the high-ionization UFO to $\sim 200$--$600\,R_\mathrm{g}$, where $R_\mathrm{g} \equiv GM_\mathrm{BH}/c^2$ and $M_\mathrm{BH}$ is the SMBH mass. In contrast, the low-ionization UFO in the soft X-ray band was not the focus of that work, and its time variability was not investigated. For the low-ionization UFO study, the same campaign has two key advantages: long X-ray monitoring over $\sim 500$~ks and simultaneous broadband coverage over 0.4--78~keV with \xrism\ and \nustar. This combination enables us to study soft-X-ray absorption variability while constraining the broadband continuum. In this paper, we use these data to study the soft X-ray variability of the low-ionization UFO in PDS~456, probe its geometry, and thereby constrain the UFO launching mechanism.}

\section{Observations \revisone{and Data Reductions}}
The main goal of this paper is to investigate low-ionization UFOs, whose absorption features mainly appear below $\sim$3~keV. \revistwo{Among the data obtained in the March 2024 campaign, we use \xrism\ \revis{\citep{XRISM_Mission2025}} and \nustar\ \revis{\citep{NuSTAR_Mission2013}} as our primary datasets. \revis{The corresponding observation IDs are 300072010 and 60901011002, respectively.}
In terms of the elapsed observing span, the \xrism\ and \nustar\ observations cover 519~ks and 325~ks, respectively. The \xmmnewton\ observation \revis{\citep{XMMNewton2001}} covers only 124~ks and is not suited to the detailed time-variability analysis pursued here. In terms of energy coverage, the combination of \xrism/\xtend\ \revis{\citep{XRISM_Xtend2025a}} and \nustar\ provides broadband spectral coverage over 0.4--78~keV. The \xrism/\xtend\ data cover 0.4--12~keV and are essential for probing the soft X-ray band below 3~keV, where signatures of low-ionization UFOs are expected to be most prominent. The \nustar\ data cover 3--78~keV and constrain the continuum above 10~keV, which is crucial for accurately modeling the broadband spectral shape.} \revisthree{For the spectral fitting, however, we exclude the \nustar\ data above 30~keV, where the background begins to dominate over the source.}
\revistwo{The \xrism/\resolve\ gate valve was closed during this observation, limiting its sensitivity in the soft X-ray band. We therefore do not use \resolve\ as a primary dataset for the low-ionization UFO analysis. Nevertheless, we make supplementary use of the high energy resolution of \resolve\ \revis{\citep{XRISM_Resolve2025}} to examine narrow absorption structures, both around the Fe-K band for the high-ionization UFO and in the 2--4~keV band.} The \swift\ data are likewise used in a supporting role to measure the UV flux and thereby constrain the spectral energy distribution (SED) of the source. The observational data described above are identical to those used in \citet{Firstpaper2025}.\par
We processed all \xrism\ data using \xrism\ \textsc{ftools} with \textsc{heasoft} ver. 6.35, and the \nustar\ data using \textsc{nustardas} v2.1.4 with \textsc{heasoft} ver. 6.35. For the \xrism/\xtend\ data, we used calibration files released on 2025-09-15. The source light curve and spectrum were extracted from a circular region with a radius of $1.'58$, while the background was extracted from a $13.'584\times 3.'168$ box region. We then generated the redistribution matrix file (\textsc{rmf}) using \textsc{xtdrmf} and the ancillary response file (\textsc{arf}) using \textsc{xaarfgen}. The \xrism/\resolve\ data were also processed with calibration files released on 2025-09-15. We excluded data from Pixel 27 because of an anomalous gain. The spectra, \textsc{rmf}, and \textsc{arf} files were produced in the same way as in \citet{Firstpaper2025}.\par
\nustar\ data were processed with calibration files released on 2024-10-01. For the FPMA detector, the source light curve and spectrum were extracted from a circular region with a radius of $89.''925$, and the background from a circular region with a radius of $211.''815$. For the FPMB detector, the source region was the same circular region of radius $89.''925$. However, because the FPMB image contains stray light in the source area, the background region was carefully chosen to be a circular region with a radius of $87.''183$ so as to include the stray-light region and minimize its effects. The source and background spectra were extracted using these same regions. Finally, the \textsc{rmf} and \textsc{arf} files were produced with \textsc{nuproducts}. We confirmed that the FPMA and FPMB data are consistent in both the spectra and the light curves. \revis{The FPMA and FPMB spectra were loaded separately and fitted simultaneously.}


\section{Analysis and Results}
\subsection{\revisone{Time variability} in soft X-rays}\label{subsec:soft_xray_time_variability}
\begin{figure}[tbp]
  \begin{center}
    \includegraphics[width=8cm]{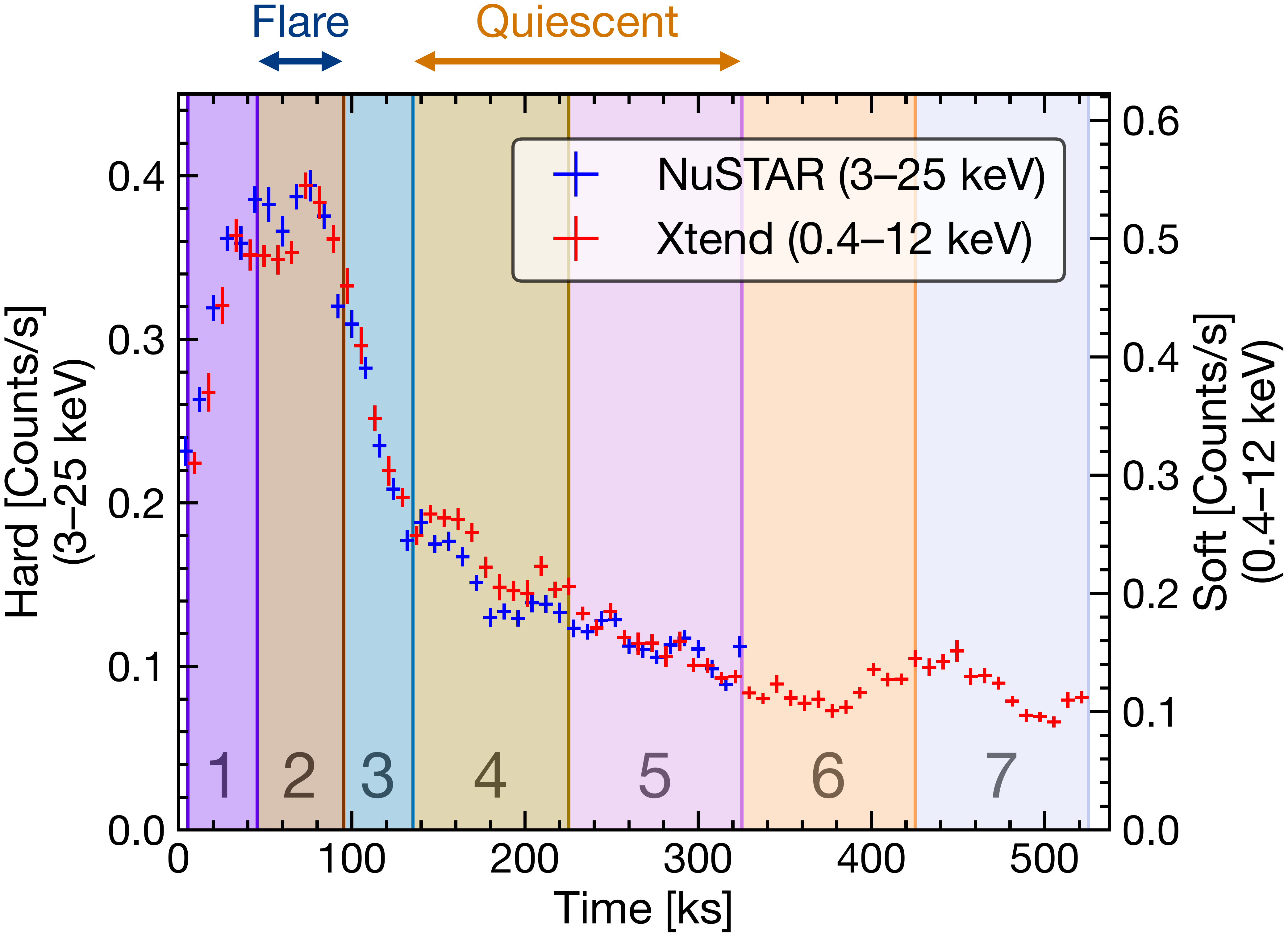}
  \end{center}
  \caption{Light curves of PDS~456 obtained with \nustar\ (3--\revis{25}~keV) and \xrism/\xtend\ (0.4--12~keV). Time is measured from the start of the \xrism\ observation. The blue and red points represent the \nustar\ and \xtend\ data, respectively. The differently colored shaded regions indicate the seven time slices used for the time-sliced spectral analysis.
  {Alt text: Line graph with time from zero to 520 kiloseconds on the x axis. The left y axis shows hard count rate from zero to 0.4 counts per second, and the right y axis shows soft count rate from zero to 0.55 counts per second. Seven shaded time intervals are labeled 1 to 7. Both light curves peak near 70 kiloseconds and decline afterward. }
  }\label{fig:7period_lightcurve}
\end{figure}
\begin{figure}[tbp]
    \begin{center}
      \includegraphics[width=8cm]{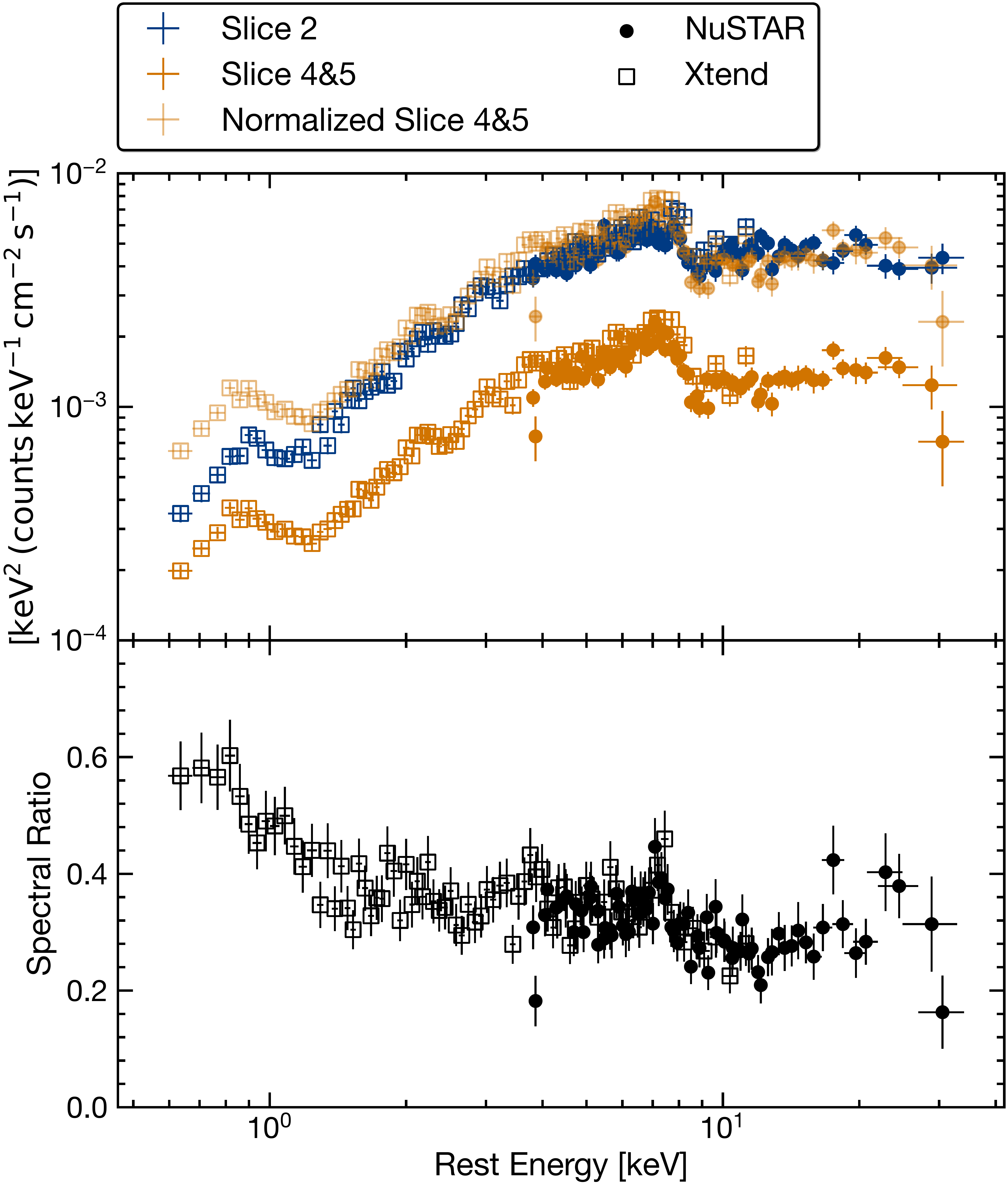}
    \end{center}
    \caption{
    \revistwo{Upper panel: Spectra in the flare phase (slice 2; blue) and quiescent phase (slices 4 and 5; orange). The light-orange points show the quiescent spectrum normalized to overlap with the flare-phase spectrum. }
    Lower panel: Quiescent-to-flare spectral ratio. Circles and squares denote the \nustar\ and \xtend\ data, respectively, and the same marker convention is used in both panels.
    {Alt text: \revistwo{Two stacked graphs. After normalization, the quiescent spectrum nearly overlaps the flare spectrum above about one kilo electron volt but remains higher at lower energies. The spectral ratio decreases from about 0.6 below one kilo electron volt to about 0.2--0.4 above several kilo electron volt. }}
    }\label{fig:spectral_ratio}
\end{figure}
To search for spectral-shape variability, we first performed a model-independent comparison between the flare and quiescent phases. Figure~\ref{fig:7period_lightcurve} shows the light curves of PDS~456 obtained with \nustar\ and \xrism/\xtend. Here, 0~ks denotes the start of the \xrism\ observation. The light curves show significant variability, including a flare-like enhancement around 80~ks. 
\revis{We divide the \xrism\ observation into seven temporal slices based on the observed flux evolution. The slice durations were chosen to follow the pre-flare, flare, and post-flare behavior, while longer intervals were generally adopted at lower count rates to ensure adequate photon statistics. Slice 1 (0--40~ks) covers the pre-flare interval, slice 2 (40--90~ks) contains the flare, and slice 3 (90--130~ks) follows the initial decline from the flare. Slices 4 (130--220~ks) and 5 (220--320~ks) sample the subsequent quiescent period, while slices 6 (320--420~ks) and 7 (420--519~ks) cover the later low-flux part of the observation. The first five slices have simultaneous \nustar\ coverage, whereas slices 6 and 7 have only \xrism\ coverage. All seven slices are used for the time-sliced spectral analysis presented in section~3.3.}
For the spectral-ratio comparison presented here, we use \revisone{slice 2 as the flare phase and combine slices 4 and 5 as the quiescent phase}. The upper panel of Fig.~\ref{fig:spectral_ratio} compares the spectra in these two phases. The spectra differ in the soft X-ray band but are similar in the hard X-ray band. This behavior is shown more clearly by the quiescent-to-flare spectral ratio in the lower panel. If the spectral shape were unchanged, this ratio would be constant over the full energy range. Instead, the ratio increases around 1~keV in the rest frame, indicating spectral variability at this energy. \revisone{A similar spectral-ratio structure was previously reported for PDS~456 and attributed to changes in the partial covering fraction of a low-ionization UFO absorber \citep{Midooka2026_lowUFO}. This similarity makes it natural to include a low-ionization UFO component in the present data and suggests that its time variability can account for the observed soft X-ray spectral changes.}
\subsection{Time-averaged spectra and broad-band modeling}
To investigate the detailed spectral variability in the soft X-ray band, \revisone{we first use the time-averaged spectra to establish the baseline spectral model and its parameters, and then apply this baseline model to the time-sliced spectra.}
This approach allows us to examine variations in the spectral parameters, particularly those associated with the low-ionization UFO components that are expected to shape the soft X-ray band. \par
For the spectral modeling, we used the \xtend, \resolve, and \nustar\ data obtained during the overlapping observation period to \revistwo{determine} the time-averaged spectral model.
\revis{In the Fe-K band (5--10~keV in the observed frame), we excluded the \nustar\ spectrum from the fit so that the absorption structure was constrained by the much higher energy resolution of \xrism/\resolve. This choice prevents the lower-resolution but higher-count-rate \nustar\ data from influencing the characterization of the resolved Fe-K absorption-line structure.}
\revis{In the broadband spectral model, we included absorption and emission components associated with both the high- and low-ionization UFOs. For the high-ionization UFO, \citet{Firstpaper2025} detected and modeled its absorption and emission features using the high-resolution \xrism/\resolve\ spectrum obtained during the same observing campaign. We therefore included the corresponding high-ionization UFO components following their model. For the low-ionization UFO, partial-covering absorption in PDS~456 was previously reported by \citet{gofford_2014_lowUFO} using \suzaku/\xis\ CCD spectra and by \citet{reeves_2016_low_UFO} using \xmmnewton/\rgs\ grating spectra. The spectral-ratio analysis in section~\ref{subsec:soft_xray_time_variability} provides further evidence for this interpretation. We therefore included a partial-covering low-ionization UFO absorption component, together with an associated emission component in the broadband model.}
\revisone{Specifically,} we define the spectral model as
\begin{multline}\label{eq:model1}
  C_{\rm cal} \cdot \mathtt{abs}_{\tt Gal}
  \times \big[ (C_\mathrm{f}\cdot \mathtt{abs}_{\tt low} \cdot \mathtt{abs}_{\tt high} + 1-C_\mathrm{f}) \cdot \mathtt{pow} \\
  + \mathtt{emiss}_{\tt high} + \mathtt{emiss}_{\tt low} \big],
\end{multline}
Here, $C_{\rm cal}$ is the cross-calibration constant between the detectors, and $\mathtt{abs}_{\tt Gal}$ represents Galactic neutral absorption modeled with \texttt{tbabs} in \revis{\textsc{xspec} \citep{XSPEC1996a}}, assuming the elemental abundances of \citet{wilms_abund}. \revistwo{The terms $\mathtt{abs}_{\tt low}$ and $\mathtt{abs}_{\tt high}$ represent absorption by the low- and high-ionization UFOs, respectively, while $\mathtt{emiss}_{\tt low}$ and $\mathtt{emiss}_{\tt high}$ represent line emission from the corresponding UFOs. All four UFO components are based on \textsc{xstar} models \revis{\citep{XSTAR2001}} calculated with the same \revis{input} SED as in \citet{Firstpaper2025}. For the absorption components, $\mathtt{abs}_{\tt high}$ includes five absorbers with different outflow velocities, as in \citet{Firstpaper2025}, while $\mathtt{abs}_{\tt low}$ is modeled with a single absorber.
For the emission components, $\mathtt{emiss}_{\tt low}$ and $\mathtt{emiss}_{\tt high}$ are convolved with an emission-line profile constructed by convolving contributions from radially outflowing hemispherical-shell profiles with outflow velocities of $\sim 0.2$--$0.3c$, following the prescription of \citet{Firstpaper2025}.}
\revis{With the outflow-induced Doppler shifts and broadening already incorporated in this profile, the XSTAR redshift parameter of $\mathtt{emiss}_{\tt low}$ was fixed at the cosmological redshift of PDS~456, $z=0.184$.}
The term $\mathtt{pow}$ denotes the power-law continuum, and $C_\mathrm{f}$ is the covering fraction of the low-ionization absorber.
\par
We note that the covering fraction formally applies to both the high- and low-ionization absorption components in equation~(\ref{eq:model1}). In practice, we interpret it as effectively representing the covering fraction of the low-ionization UFO. This is because the spectral signature of the high-ionization UFO is mainly confined to the Fe-K band around 7--8~keV in the rest frame, and its covering fraction cannot be constrained independently of the column density. Therefore, in equation~(\ref{eq:model1}), the covering-fraction parameter primarily reflects the covering fraction of the low-ionization UFO component. The parameters of the high-ionization UFO components ($\mathtt{abs}_{\tt high}$ and $\mathtt{emiss}_{\tt high}$), except for the column density of $\mathtt{abs}_{\tt high}$, were fixed to the best-fit values derived by \citet{Firstpaper2025}, who analyzed the same dataset. \revis{The ionization parameter was fixed at $\log(\xi/(\unit{\erg\cm\per\s}))=4.90$, and the turbulent velocity of $\mathtt{abs}_{\tt high}$ was fixed at $v_\mathrm{turb}=1900~\mathrm{km~s^{-1}}$. The five outflow velocities of $\mathtt{abs}_{\tt high}$ were fixed at $v_\mathrm{out}/c=0.226$, 0.254, 0.278, 0.307, and 0.333. The normalization of $\mathtt{emiss}_{\tt high}$ was fixed at $\kappa=\num{2.1e-4}$, corresponding to a global covering factor of $f_\mathrm{cov}=1.9$ according to equation~(4) of \citet{Firstpaper2025}.} We allowed the column density of $\mathtt{abs}_{\tt high}$ to vary freely so that the covering fraction would not be driven by the Fe-K absorption associated with the high-ionization UFO. \revis{The spectral fitting was performed with \textsc{xspec} using Cash statistics \revis{\citep{Cash1979}}.} The \revistwo{90\%} errors were calculated using the \textsc{error} command in \textsc{xspec}, which computes the confidence interval where the C-statistic increases by \revistwo{2.706} from its minimum for a single parameter of interest.
\begin{figure}[tbp]
  \begin{center}
    \includegraphics[width=8cm]{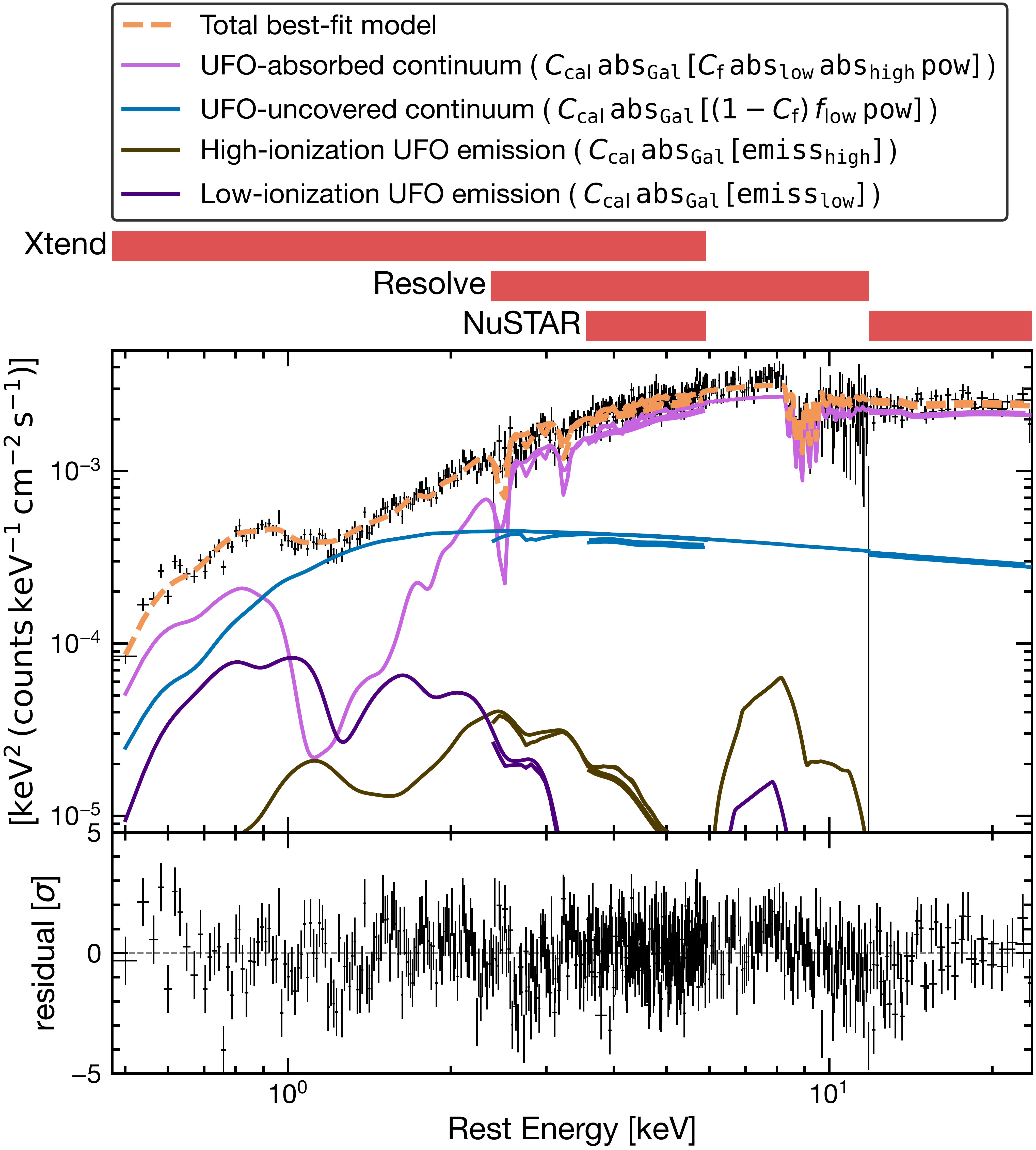}
  \end{center}
  \caption{Time-averaged spectra overlaid with the best-fit broadband model. \revisone{Red bars indicate the fitting energy ranges for \xtend, \resolve, and \nustar.} Black points show the observed data from \revisone{these instruments}. \revisone{Dashed lines show the total best-fit model, and solid lines show individual model components. Both are shown separately for each detector.} The lower panel shows the residuals, defined as $\mathrm{(data - model) / error}$. {Alt text: Two stacked graphs \revisone{with red horizontal bars above the spectra labeled Xtend, Resolve, and NuSTAR.} The upper graph shows spectra and model curves versus rest energy from about 0.5 to 25 kilo electron volt on logarithmic axes. The lower graph shows residuals mostly distributed around zero.}
  }\label{fig:wholetime-spectrum1}
\end{figure}
\begin{table}[tbp]
  \tbl{Time-averaged spectral fitting results for the low-ionization UFO components.}{
\begin{tabular}{llc}
\hline
Component & Parameter & Value \\
\hline
\multirow{5}{*}[-5pt]{$\mathtt{abs_{low}}$} & $N_\mathrm{H}$\textsuperscript{a} & $\num{8.00(9:10)}$ \rule[-6pt]{0pt}{16pt} \\
 & $\log \xi$\textsuperscript{b} & $\num{3.117(8)}$ \rule[-6pt]{0pt}{16pt}\\
 & $v_\mathrm{turb}$\textsuperscript{c} & $\num{1.24(9:8)e3}$ \rule[-6pt]{0pt}{16pt} \\
 & $v_\mathrm{out}$\textsuperscript{d} & $\num{0.2940(28:26)}$ \rule[-6pt]{0pt}{16pt} \\
\cline{1-3}
\multirow{1}{*}{\revistwo{$C_\mathrm{f}$}} & & $\num{0.8885(19:21)}$  \rule[-6pt]{0pt}{16pt} \\
\cline{1-3}
  \multirow{3}{*}[-6pt]{$\mathtt{emiss_{low}}$} & $N_\mathrm{H}$\textsuperscript{a} & $\num{8.00}^\dagger$ \rule[-6pt]{0pt}{16pt} \\
 & $\log \xi$\textsuperscript{b} & $\num{3.117}^\dagger$\rule[-6pt]{0pt}{16pt} \\
 & norm\textsuperscript{e}  & $\num{1.22(12:13)e-4}$ \rule[-6pt]{0pt}{16pt} \\ \hline
 $C/\text{d.o.f.}$ & \multicolumn{2}{c}{$1.02\,(17741/17344)$} \rule[-2pt]{0pt}{11pt} \\
\hline
\end{tabular}
}\label{tab:wholetime_fitting}
\begin{tabnote}
  \par\noindent
  \hbox to6pt{$^{\rm a}$\hss}
  \unskip Column density in units of $10^{22}~\mathrm{cm^{-2}}$.
  \par\noindent
  \hbox to6pt{$^{\rm b}$\hss}
  \unskip Ionization parameter in units of $\mathrm{erg~cm~s^{-1}}$.
  \par\noindent
  \hbox to6pt{$^{\rm c}$\hss}
  \unskip Turbulence velocity in units of $\mathrm{km~s^{-1}}$.
  \par\noindent
  \hbox to6pt{$^{\rm d}$\hss}
  \unskip Outflow velocity in units of the speed of light $c$.
  \par\noindent
  \hbox to6pt{$^{\rm e}$\hss}
  \unskip \revisone{The normalization $\kappa$ scales with the outflow covering factor $f_\mathrm{cov} \equiv \Omega/2\pi$, where $\Omega$ is the near-side wind solid angle. See equation~(4) of \citet{Firstpaper2025} for the definition.}
  \par\noindent
  \hbox to6pt{$^\dagger$\hss}
  \unskip \revisone{Linked to the corresponding absorption-component parameters.}
  \end{tabnote}
\end{table}
\par
\begin{figure}[tbp]
  \begin{center}
    \includegraphics[width=8cm]{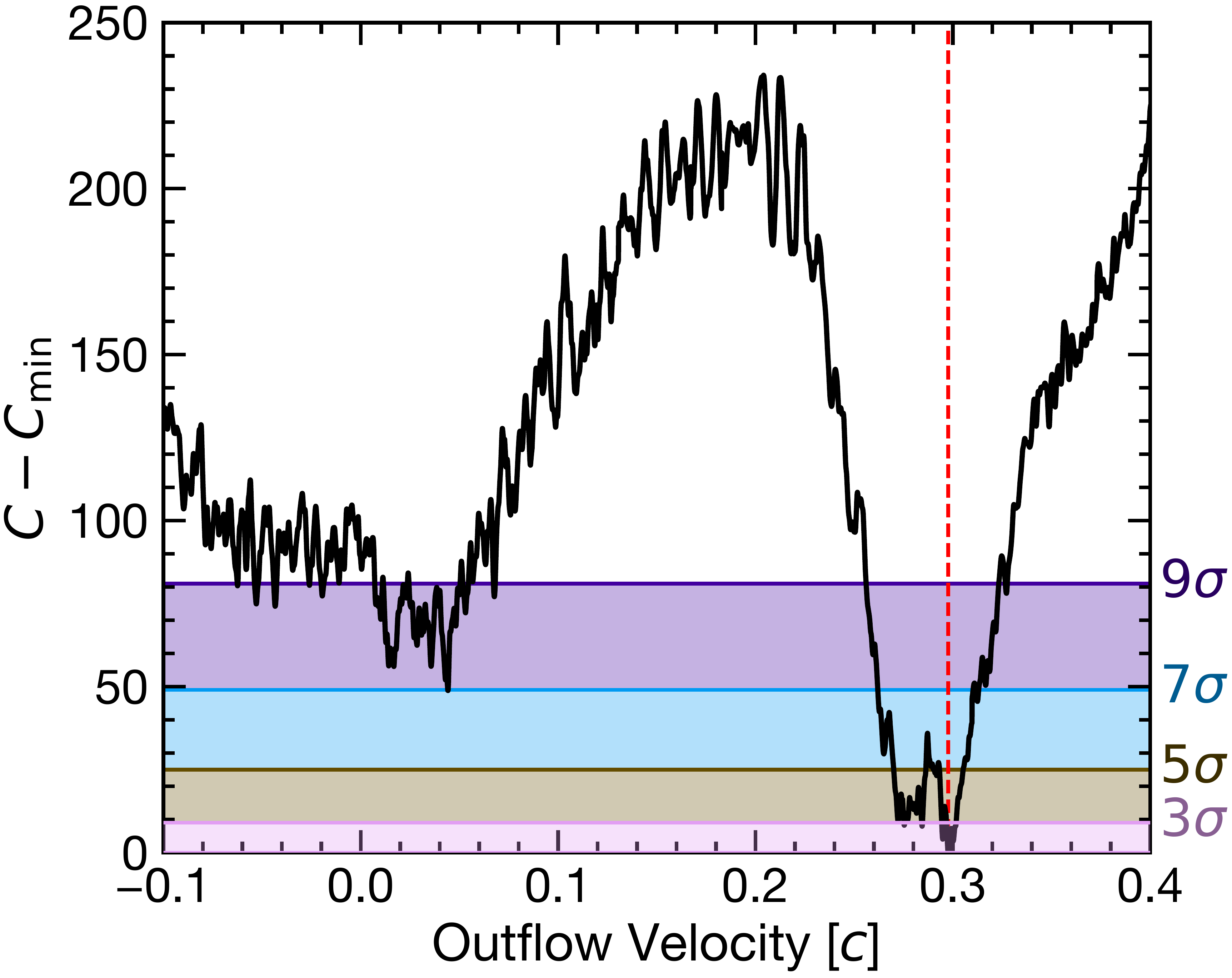}
  \end{center}
  \caption{
    C-statistic relative to the minimum value, $C-C_\mathrm{min}$, as a function of the outflow velocity of the low-ionization absorber. The shaded regions represent the $3\sigma$, $5\sigma$, $7\sigma$, and $9\sigma$ levels. The red dashed line indicates the best-fit outflow velocity of the low-ionization UFO. {Alt text: Line graph with outflow velocity from minus 0.1 to 0.4 times the speed of light on the x axis and C minus C minimum from about zero to one hundred on a linear y axis. The curve has a deep minimum near 0.30 times the speed of light. Compared with the rest-frame absorber at zero velocity, the decrease in C-statistic exceeds the seven sigma level. }
  }\label{fig:velocity_vs_cstat}
\end{figure}

Figure~\ref{fig:wholetime-spectrum1} shows the best-fit model, and table~\ref{tab:wholetime_fitting} summarizes the best-fit parameters for the low-ionization UFO components. The broadband spectra are successfully explained by the emission and absorption components of the high- and low-ionization UFOs, and the partial-covering absorption features of the low-ionization UFO are prominent around 1--2~keV. \revisone{The tight constraint on the outflow velocity listed in table~\ref{tab:wholetime_fitting} is mainly \revistwo{driven by the soft-X-ray \resolve\ data around 2--4~keV; in particular, absorption-line structures from Si and S ions help determine the velocity.}}
Figure~\ref{fig:velocity_vs_cstat} shows $C-C_\mathrm{min}$ as a function of the outflow velocity of the low-ionization absorber, which we used to evaluate the statistical significance of the low-ionization UFO detection. The minimum C-statistic is obtained at an outflow velocity of $\sim 0.3c$. \revistwo{Relative to a stationary absorber with $v_\mathrm{out}=0$,} the decrease in the C-statistic is $C_\mathrm{rest}-C_\mathrm{min}\sim 50$, indicating a highly significant detection of the low-ionization UFO at this velocity, with a significance exceeding $7\sigma$ \revistwo{according to Wilks' theorem \revis{\citep{Wilks1938}}}. \revistwo{The curve shows two troughs: although table~\ref{tab:wholetime_fitting} gives a tightly constrained best-fit velocity, a slower local minimum remains possible at about the $3\sigma$ level, and the conservative $5\sigma$ confidence range is $v_\mathrm{out}\simeq 0.26$--$0.31c$.}
\subsection{Time-sliced spectra}
We investigated the time-sliced spectra to characterize the soft X-ray variability associated with the low-ionization UFO. Figure~\ref{fig:7period_lightcurve} defines the seven time slices into which we divided the entire \xrism/\xtend\ observation. For slices 1--5, simultaneous \nustar\ observations were also available, allowing us to perform broadband spectral fitting for these slices. For slices 6 and 7, we used only the \xrism/\xtend\ data for the spectral fitting. \par
For the time-sliced spectral fitting, we used the same model (equation \ref{eq:model1}) as in the time-averaged analysis. The aim of this analysis is to characterize the variability seen in the soft X-ray band, where the low-ionization UFO primarily shapes the spectrum.
\revis{For the Fe-K band, all parameters of $\mathtt{emiss}_{\tt high}$ were fixed to their time-averaged values, consistent with the lack of significant emission variability reported by \citet{Firstpaper2025}. Because the time-sliced fits do not include \resolve, they cannot independently constrain the kinematics or relative strengths of the five components in $\mathtt{abs}_{\tt high}$, although they remain sensitive to changes in the overall Fe-K absorption strength. We therefore fixed the ionization parameter, turbulent velocity, five outflow velocities, and relative column-density ratios to their time-averaged values, and allowed only the total column density to vary independently among the time slices as a single measure of the overall absorption-strength variations.}
\revis{For the continuum, we linked the photon index across all time slices because the spectral shape above 10~keV shows no significant variability (Fig.~\ref{fig:spectral_ratio}; see also Appendix~\ref{appendix:continuum_variability}).} \revistwo{The power-law normalization, however, was allowed to vary independently among the seven slices.}
\revis{For the low-ionization UFO,} we linked the ionization parameter, turbulent velocity, and outflow velocity across all time slices, assuming that these parameters do not vary on the timescale of the observation. \revistwomaj{The decision to link the ionization parameter and outflow velocity was supported by an additional fit in which these parameters were allowed to vary independently among the time slices (see Appendix~\ref{appendix:time_sliced_fitting_strategy}).}\par
\begin{figure*}[tbp]
  \begin{center}
    \includegraphics[width=17.8cm]{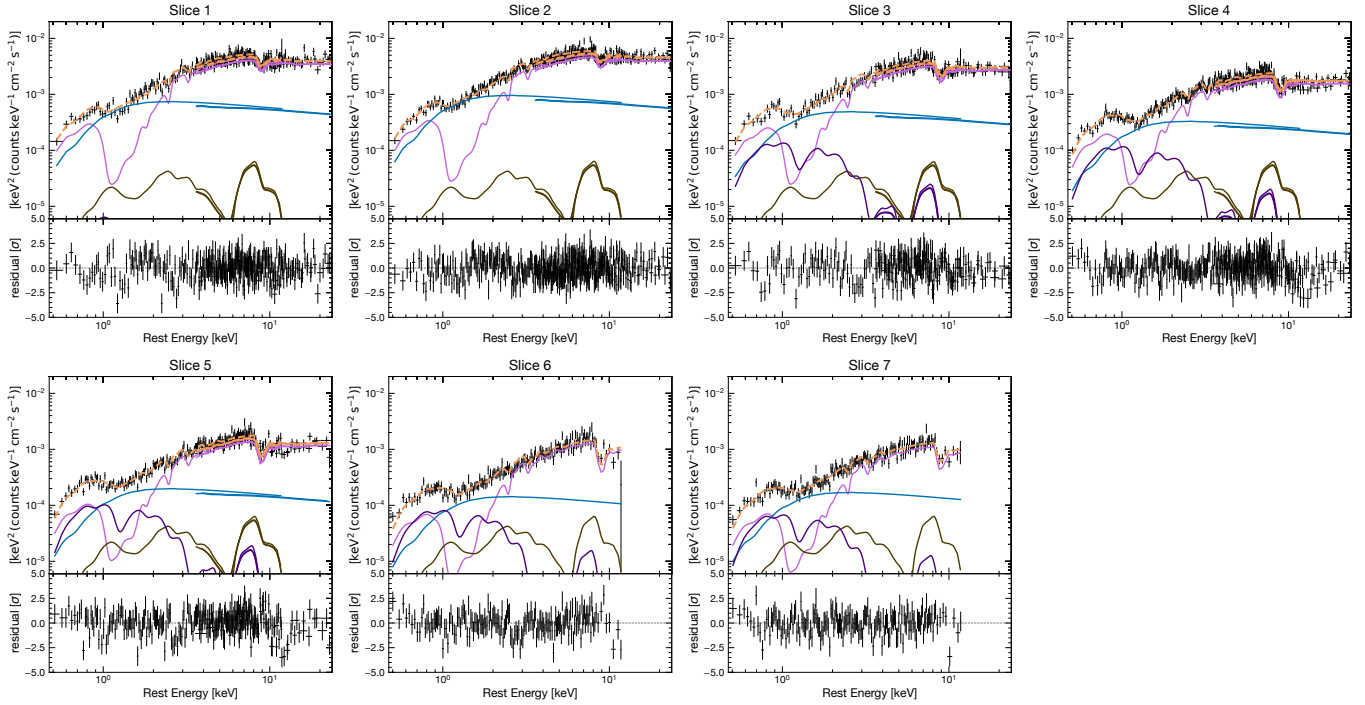}
  \end{center}
  \caption{Seven time-sliced spectra overlaid with the best-fit models, along with residual plots. The upper panel of each slice shows the spectra, while the lower panel shows the residuals. The black points represent the observed data from \nustar\ and \xrism/\xtend. The dashed lines represent the best-fit models, and the solid lines show their individual components. The color coding of the lines is the same as in Fig.~\ref{fig:wholetime-spectrum1}. {Alt text: Seven panels arranged as four on top and three below. Each panel has an upper spectrum and a lower residual graph versus rest energy from about 0.5 to 25 kilo electron volt. The residual axes span minus five to five standard deviations, but most residual points are clustered within about plus or minus 2.5 standard deviations. }
  }\label{fig:7period_spectrum}
  \label{fig:wide}
\end{figure*}

\begin{figure}[tbp]
  \begin{center}
    \includegraphics[width=8cm]{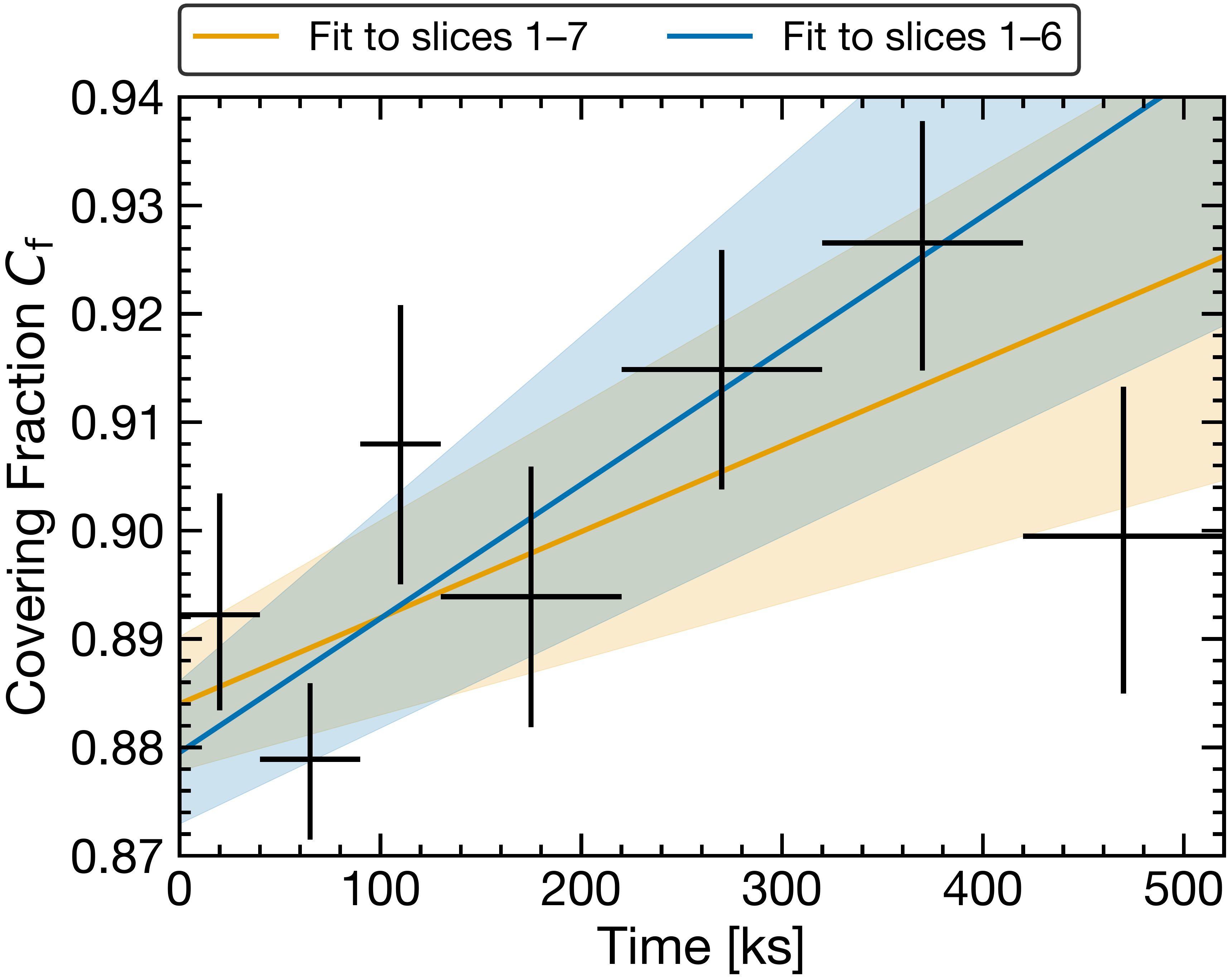}
  \end{center}
  \caption{Variation in the covering fraction of the low-ionization UFO across the seven time slices. The horizontal error bars represent the duration of each time slice, while the vertical error bars represent the $1\sigma$ confidence intervals of the covering fraction. The lines and shaded regions represent the best-fit linear function and its $1\sigma$ confidence interval, respectively.
  {Alt text: Line graph with time from zero to 500 kiloseconds on the x axis and covering fraction from 0.87 to 0.94 on the y axis. Seven points with error bars generally increase from about 0.89 to 0.92, while the last point near 400 kiloseconds drops below the increasing trend. }
  }\label{fig:7period_covfrac_var}
\end{figure}

 
Figure~\ref{fig:7period_spectrum} shows the spectra of the seven time slices overlaid with the best-fit models, and table~\ref{tab:7term_fitting} summarizes the best-fit parameters for the low-ionization UFO components. The spectral fitting successfully reproduces the spectra of all time slices, and the partial-covering absorption features of the low-ionization UFO are prominent around 1--2~keV in all slices.
\revistwo{As shown in table~\ref{tab:7term_fitting}, several model parameters vary over the course of the observation. The variability of $\mathtt{abs}_{\tt high}$ and $\mathtt{emiss}_{\tt low}$ is discussed separately in Appendices~\ref{appendix:high_ionization_UFO_variability} and \ref{appendix:low_ionization_emission}, respectively.}
\revis{This work assumes that the intrinsic X-ray spectrum is a single power law, but there could also be a contribution from a variable soft X-ray excess  at low energies and this could have an impact on the derived variability of the ionised emission components (Appendix~\ref{appendix:low_ionization_emission}). We will explore this possibility in a subsequent paper.}
\revistwo{In the following, we focus on $C_\mathrm{f}$.}
Figure~\ref{fig:7period_covfrac_var} shows the covering fraction of the low-ionization UFO obtained from the time-sliced spectral fitting.
\revistwo{A test of the time-independent covering-fraction hypothesis gives a p-value of $\num{8.5e-8}$, indicating significant covering-fraction variability.}
The covering fraction varies across the time slices and shows a mostly increasing trend, except in the last time slice. To characterize this variation, we fitted the covering fraction with a linear function. The slope is \qty{7.9(17)e-5}{\per\kilo\s} when all time slices are included, and \qty{1.3(2)e-4}{\per\kilo\s} when the last time slice is excluded. The difference between these two slopes reflects the systematic uncertainty associated with the physical interpretation of the last time slice\revisone{, as discussed in Sec.~\ref{subsubsec:covering_fraction_variability}.}
{\setlength{\tabcolsep}{4pt}
\begin{longtable}[tbp]{llccccccc}
  \caption{Time-sliced spectral fitting results for the low-ionization UFO components.}\label{tab:7term_fitting}
\hline\noalign{\vskip3pt}
\endhead
\hline\noalign{\vskip3pt}
\endfoot
\hline\noalign{\vskip3pt}
\multicolumn{9}{@{}l@{}}{\hbox to0pt{\parbox{160mm}{
  \footnotesize
  \hangindent6pt\noindent
  \hbox to6pt{$^{\rm a}$\hss}
  \unskip Power-law flux over 10--20~keV in units of $10^{-12}\,\mathrm{erg~cm^{-2}~s^{-1}}$.
  \par\noindent
  \hbox to6pt{$^{\rm b}$\hss}
  \unskip Sum of the column densities of the five high-ionization UFO clumps, in units of $10^{22}~\mathrm{cm^{-2}}$.
  \par\noindent
  \hbox to6pt{$^{\rm c}$\hss}
  \unskip Column density in units of $10^{22}~\mathrm{cm^{-2}}$.
  \par\noindent
  \hbox to6pt{$^{\rm d}$\hss}
  \unskip Ionization parameter in units of $\mathrm{erg~cm~s^{-1}}$.
  \par\noindent
  \hbox to6pt{$^{\rm e}$\hss}
  \unskip Turbulence velocity in units of $\mathrm{km~s^{-1}}$.
  \par\noindent
  \hbox to6pt{$^{\rm f}$\hss}
  \unskip Outflow velocity in units of the speed of light $c$.
  \par\noindent
  \hbox to6pt{$^{\rm g}$\hss}
  \unskip Normalization in units of $10^{-4}$.
  \par\noindent
  \hbox to6pt{}\unskip The covering fraction is calculated as $C_\mathrm{f}=1/(1+f_\mathrm{low})$.
  \par\noindent
  \hbox to6pt{$^\dagger$\hss}
  \unskip The parameters of the emission component are linked to those of the absorption component.
  \par\noindent
  \hbox to6pt{$^\ddagger$\hss}
  \unskip 90\% upper limit.
}\hss}
}
\endlastfoot
Component & Parameter & Slice 1 & Slice 2 & Slice 3 & Slice 4 & Slice 5 & Slice 6 & Slice 7 \\
\hline
 \revis{\multirow{2}{*}[-2pt]{$\mathtt{pow}$}} & \revis{photon index} & \multicolumn{7}{c}{\revis{$\num{2.256(25:23)}$}} \rule[-6pt]{0pt}{16pt} \\
 & norm\textsuperscript{a} & $\num{3.82(14:13)}$ & $\num{4.32(13)}$ & $\num{3.00(13:12)}$ & $\num{1.73(7:6)}$ & $\num{1.32(6)}$ & $\num{1.12(11:9)}$ & $\num{0.94(8)}$ \rule[-6pt]{0pt}{16pt} \\\hline
 $\mathtt{abs_{high}}$ & total $N_\mathrm{H}$\textsuperscript{b} & $\num{23(6:5)}$ & $\num{21(4)}$ & $\num{34(8:7)}$ & $\num{39(8:6)}$ & $\num{50(10)}$ & $\num{60(23:19)}$ & $\num{45(17:16)}$ \rule[-6pt]{0pt}{16pt} \\\hline
 \multirow{5}{*}[-2pt]{$\mathtt{abs_{low}}$} & $N_\mathrm{H}$\textsuperscript{c} & $\num{8.79(71:46)}$ & $\num{8.78(47)}$ & $\num{8.39(62:59)}$ & $\num{7.27(59:48)}$ & $\num{8.41(61:60)}$ & $\num{9.1(10:7)}$ & $\num{8.63(75:73)}$ \rule[-6pt]{0pt}{16pt} \\
 & $\log \xi$\textsuperscript{d} & \multicolumn{7}{c}{$\num{3.118(24:39)}$} \rule[-6pt]{0pt}{16pt} \\
 & $v_\mathrm{turb}$\textsuperscript{e} & \multicolumn{7}{c}{$\num{1.18(28:36)e3}$} \rule[-6pt]{0pt}{16pt} \\
 & $v_\mathrm{out}$\textsuperscript{f} & \multicolumn{7}{c}{$\num{0.2970(35:42)}$} \rule[-6pt]{0pt}{16pt} \\
\hline
 $C_\mathrm{f}$ & & $\num{0.892(11:9)}$ & $\num{0.879(7)}$ & $\num{0.908(13)}$ & $\num{0.894(12)}$ & $\num{0.915(11)}$ & $\num{0.927(11:12)}$ & $\num{0.899(14:15)}$ \rule[-6pt]{0pt}{16pt} \\
\hline
 \multirow{3}{*}[-5pt]{$\mathtt{emiss_{low}}$} & $N_\mathrm{H}$\textsuperscript{c} & $\num{8.8}^\dagger$ & $\num{8.8}^\dagger$ & $\num{8.4}^\dagger$ & $\num{7.3}^\dagger$ & $\num{8.4}^\dagger$ & $\num{9.1}^\dagger$ & $\num{8.6}^\dagger$ \rule[-6pt]{0pt}{16pt} \\
  & $\log \xi$\textsuperscript{d} & \multicolumn{7}{c}{$3.118^\dagger$} \rule[-6pt]{0pt}{16pt} \\
  & norm\textsuperscript{g} & $<\num{0.9}^\ddagger$ & $<\num{0.3}^\ddagger$ & $\num{1.88(75:88)}$ & $\num{1.67(43:50)}$ & $\num{1.41(33:35)}$ & $\num{1.10(29:31)}$ & $\num{0.93(29)}$ \rule[-6pt]{0pt}{16pt} \\\hline
 $C$-statistics & & 2541 & 2519 & 2672 & 2552 & 2729 & 1642 & 1667 \rule[-2pt]{0pt}{11pt}\\\hline
 bins & & 2447 & 2447 & 2447 & 2447 & 2447 & 1599 & 1599 \rule[-2pt]{0pt}{11pt}\\\hline
 $C/\text{d.o.f}$ & \multicolumn{8}{c}{$1.06\,(16323/15391)$}\rule[-2pt]{0pt}{11pt}
\end{longtable}
}

\section{Discussion}
\subsection{Distance between the low-ionization UFO and the SMBH}
\subsubsection{Covering-fraction variability and crossing velocity}
\label{subsubsec:covering_fraction_variability}
\revistwo{We interpret the covering-fraction variability as transverse motion of the absorber across the X-ray source (corona). Under this interpretation,} the increasing trend in the covering fraction suggests that the low-ionization UFO is moving into the line of sight across the X-ray source. The last time slice (slice 7) appears to deviate from this increasing trend, which may reflect a statistical fluctuation, the low-ionization UFO moving out of the line of sight, or substructure within the low-ionization UFO, such as a hole. \par
This variation is also consistent with earlier observations of the low-ionization UFO in PDS~456 \citep {Reeves2020_lowUFO}.
They reported that the column density of the low-ionization UFO changed from $N_\mathrm{H}=\qty{e23}{\cm^{-2}}$ to $N_\mathrm{H}=\qty{e21}{\cm^{-2}}$ between two observations separated by 20~days.
They interpreted this as a dense inner absorber surrounded by a more diffuse outer absorber, with the dense component having moved out of the line of sight between the two epochs. \revis{In the scenario proposed by \citet{Reeves2020_lowUFO},} the covering fraction of the dense absorber decreased from $\sim 1$ to $\sim 0$ in less than 20~days, giving $\Delta C_\mathrm{f}/\Delta t \geq \qty{6e-4}{\per\kilo\s}$. This is within the same order of magnitude as the slope range inferred from our time-sliced analysis, $\Delta C_\mathrm{f}/\Delta t \sim (0.79$--$1.3)\times 10^{-4}\,\unit{\per\kilo\s}$. Although the corona size and the distance of the low-ionization UFO from the SMBH may differ between the two epochs, and hence the crossing velocities need not be identical, this agreement in order of magnitude indicates that the two observations are mutually consistent.
\par
The covering fraction $C_\mathrm{f}$ represents the ratio of the projected area of the absorber to that of the corona. \revistwo{Under the transverse-motion interpretation,} the covering-fraction variability is related to the absorber crossing velocity across the corona, $v_\mathrm{cross}$, and the corona size, $D_\mathrm{corona}$, as $\Delta C_\mathrm{f} \sim {D_\mathrm{corona}v_\mathrm{cross}\Delta t}/{D_\mathrm{corona}^2}$.
As shown in Fig.~\ref{fig:7period_lightcurve} and discussed in \citet{Firstpaper2025}, the power-law continuum varies on a timescale of $\Delta t_\mathrm{corona}\sim 40$~ks, giving a light-crossing size of $D_\mathrm{corona} \lesssim c\Delta t_\mathrm{corona}$. Therefore, we can constrain the crossing velocity as
\begin{equation}\label{eq:vcross_upper_limit}
  v_\mathrm{cross} \sim \frac{\Delta C_\mathrm{f}}{\Delta t} D_\mathrm{corona} \lesssim \frac{\Delta C_\mathrm{f}}{\Delta t}\Delta t_\mathrm{corona} c\lesssim \num{5e-3}c.
\end{equation}
\revistwomaj{We note that internal structural changes, such as the spreading or clustering of dense fragments within the low-ionization UFO, may also contribute to the covering-fraction variability. In particular, using the same observing campaign analyzed here, \citet{Xu2025} found that the low-ionization UFO in PDS~456 lies in a thermally unstable region, providing a possible physical origin for such structural changes. In this case, the observed variability need not be described as a steady increase in covering fraction, and the value of $\Delta C_\mathrm{f}/\Delta t$ adopted above need not represent transverse motion alone. If part or all of the observed variability is attributed to such changes, the crossing velocity required to account for the remaining variability becomes smaller, or even zero. Thus, the upper-bound inequality, $v_\mathrm{cross}\lesssim \num{5e-3}c$, in equation~(\ref{eq:vcross_upper_limit}) remains applicable.}
\subsubsection{Estimation of the distance}
\label{subsubsec:distance_estimation}
We can then estimate the distance of the low-ionization UFO from the SMBH using the crossing velocity. We apply conservation of the specific angular momentum component parallel to the disk rotation axis between the launching point and the observed location of the UFO.
Conservation of the axial specific angular momentum can be written as:
\begin{equation}
  R v_{\phi} = R_0 v_{0\phi} + \Delta L_\mathrm{mag}.
\end{equation}
Here, $v_{0\phi}$ and $v_\phi$ are the azimuthal (disk-rotation) velocity components at the launching point and at the observed location, respectively; $R_0$ and $r$ are the distances from the SMBH to these two locations; $R \equiv r\sin\theta_i$ is the projection of $r$ onto the disk plane for the inclination angle $\theta_i$; and $\Delta L_\mathrm{mag}$ represents the specific angular momentum transferred from the accretion disk to the low-ionization UFO via magnetic field lines. \revisone{\revistwo{If the outflow is driven by radiation pressure,} then $\Delta L_\mathrm{mag} \approx 0$, because the radiative force is predominantly radial and does not efficiently transfer angular momentum.} \revisone{On the other hand, \revistwo{if the outflow is magnetically driven,} the outflow receives angular momentum from the disk through magnetic stresses, giving $\Delta L_\mathrm{mag} > 0$ \citep{Blandford1982}.}
In either case, we have $R v_{\phi} \geq R_0 v_{0\phi}$.
To constrain the distance $R$ from this inequality, \revisone{we take $v_{0\phi}$ to be the Keplerian velocity at $R_0$}, i.e., $v_{0\phi} = c\sqrt{R_\mathrm{g}/R_0}$.
\revisone{The crossing velocity represents the transverse motion of the absorber across the corona and may include not only the rotational motion but also a projected component of the outflow motion. Therefore, the azimuthal component satisfies $v_\phi \leq v_\mathrm{cross}$.} Combining this with $Rv_\phi \geq R_0 v_{0\phi} = c\sqrt{R_\mathrm{g} R_0}$, we obtain the following lower limit on the distance:
\begin{equation}
  r = \frac{R}{\sin\theta_i} \geq \frac{R_\mathrm{g}}{\sin\theta_i}
      \left(\frac{v_\mathrm{cross}}{c}\right)^{-1}
      \left(\frac{R_0}{R_\mathrm{g}}\right)^{1/2}.
  \label{eq:rdist_lower}
\end{equation}
The launching radius $R_0$ is bounded below by the escape radius, $R_\mathrm{esc} \equiv 2R_\mathrm{g}(c/v_\mathrm{out})^2$, at which the outflow velocity equals the local escape velocity. For $v_\mathrm{out} \sim 0.3c$, this gives $R_0 \gtrsim R_\mathrm{esc} \approx 22\,R_\mathrm{g}$. Substituting $R_0 \gtrsim 22\,R_\mathrm{g}$ and $v_\mathrm{cross} \lesssim \num{5e-3}c$ into Equation~(\ref{eq:rdist_lower}), \revis{we adopt $\theta_i \sim 13^\circ$, the latest estimate of the nuclear broad-line region inclination obtained by \citet{Amorim2024} through dynamical modeling of spatially resolved Pa$\alpha$ emission observed with VLTI/GRAVITY. With this value, we obtain $r \gtrsim \num{4e3}\,R_\mathrm{g}$.}
This suggests that the low-ionization UFO is located at a distance of at least a few thousand gravitational radii from the SMBH. \revistwomaj{This lower limit does not conflict with previous estimates for the soft-X-ray UFO in PDS~456; in particular, \citet{Reeves2020_lowUFO} suggested a parsec-scale location, about 1~pc, corresponding to $\sim4\times10^4\,R_\mathrm{g}$ for $M_\mathrm{BH}\sim\num{5e8}\,M_\odot$ \citep{GravityCollab_2023}.} \par
\revistwomaj{
  \revis{An independent estimate can also be obtained from the size constraint and ionization parameter of the low-ionization UFO. Its fitted covering fraction, $C_\mathrm{f}=0.88$--$0.93$, indicates that it covers most of the X-ray source. Its projected size, $D_\mathrm{abs}$, should therefore be at least comparable to the coronal size,} $D_\mathrm{abs}\gtrsim D_\mathrm{corona}\sim c\Delta t_\mathrm{corona}\simeq \qty{1.2e15}{\cm}$ for $\Delta t_\mathrm{corona}\sim 40$~ks. With the fitted column density $N_\mathrm{H}\sim\qty{8e22}{\cm^{-2}}$, this gives the number density $n\sim N_\mathrm{H}/D_\mathrm{abs}\lesssim N_\mathrm{H}/D_\mathrm{corona}\sim\qty{6e7}{\cm^{-3}}$. Assuming negligible shielding, we take the ionizing luminosity seen by the low-ionization UFO to be the value inferred by \citet{Firstpaper2025}, $L_\mathrm{ion}\simeq \qty{1.6e46}{\erg\per\s}$. Combining this with $\xi=L_\mathrm{ion}/(nr^2)$ and $\log\xi\simeq 3.1$ gives $r\gtrsim\qty{5e17}{\cm}\sim\num{7e3}\,R_\mathrm{g}$ for $M_\mathrm{BH}\sim\num{5e8}\,M_\odot$ \citep{GravityCollab_2023}. Thus, this independent estimate agrees with the crossing-velocity-based lower limit above to \revisthree{an} order of magnitude.
}\par
\revistwomaj{The inferred crossing velocity can also be checked against the measured turbulent velocity by requiring that the associated line-of-sight velocity broadening not exceed the observed turbulent line width. Following the geometry illustrated in Fig.~1 of \citet{FukumuraTombesi2019}, we consider an almost fully covering absorber moving transversely in front of a spherical corona of diameter $D_\mathrm{corona}$. Because the azimuthal-velocity directions sampled against the two sides of the corona differ, their line-of-sight projections broaden the absorption line. Adopting the same angular definitions as \citet{FukumuraTombesi2019} and identifying their coronal radius with $D_\mathrm{corona}/2$, $\phi$ is the azimuthal angle of a sightline through the corona measured from the sightline through its center, and $\alpha\simeq D_\mathrm{corona}/(2r)$ is the azimuthal half-angle subtended by the corona at the absorber. The line-of-sight velocity shift due to the azimuthal motion is then $\Delta v_\mathrm{los}(\phi)\equiv v_\phi\sin\theta_i\sin\phi$. Defining $\Delta v_\mathrm{rms}$ as the root-mean-square of $\Delta v_\mathrm{los}(\phi)$ over the corona, we obtain $\Delta v_\mathrm{rms}\simeq v_\phi\sin\theta_i\,\alpha/\sqrt{3}\leq v_\mathrm{cross}\sin\theta_i D_\mathrm{corona}/(2\sqrt{3}r)$ for $\alpha\ll1$, where we used $v_\phi\leq v_\mathrm{cross}$. Using $v_\mathrm{cross}\lesssim\num{5e-3}c$, $\theta_i\simeq13^\circ$, $D_\mathrm{corona}\lesssim c\Delta t_\mathrm{corona}\simeq16\,R_\mathrm{g}$ for $\Delta t_\mathrm{corona}\sim40$~ks, and $r\gtrsim\num{4e3}\,R_\mathrm{g}$ gives $\Delta v_\mathrm{rms}\lesssim\qty{0.4}{\km\per\s}$. This is much smaller than the measured turbulent velocity of $\sim1.2\times10^3\,\mathrm{km~s^{-1}}$; thus, the inferred crossing motion does not overbroaden the absorption line and is not in conflict with the measured line width. Conversely, the crossing motion contributes negligibly to the measured turbulent velocity, which must therefore be dominated by other broadening mechanisms.}

\subsubsection{Comparison with the High-Ionization UFO}\label{subsubsec:comparison_high_low}
\begin{figure}[tbp]
  \begin{center}
    \includegraphics[width=8cm]{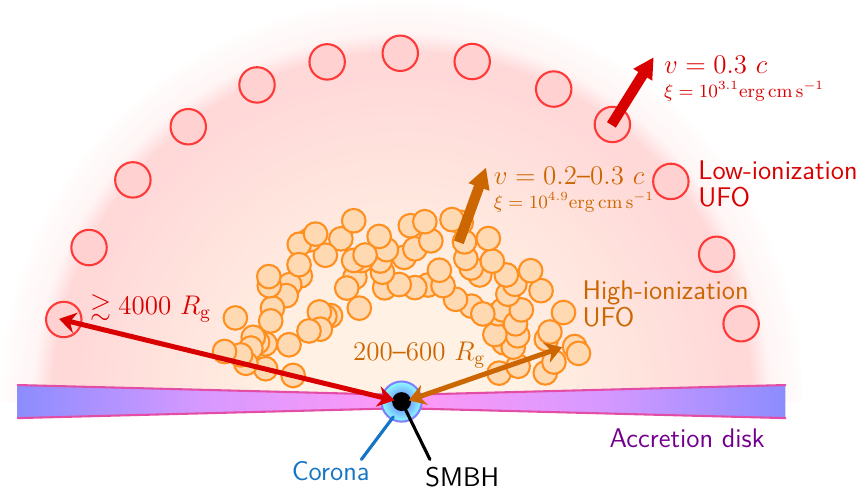}
  \end{center}
  \caption{Schematic illustration of the structural picture of the UFOs in PDS~456. The high-ionization UFO is located closer to the SMBH, while the low-ionization UFO is located farther away. Both components have similar outflow velocities, but different ionization states and distances from the SMBH; their number densities and clump sizes can also be comparable if the low-ionization UFO lies near its lower-limit distance.
  {Alt text: Schematic diagram. A supermassive black hole and corona lie at the center of an accretion disk. The high ionization ultra fast outflow forms an inner clumpy arc at 200 to 600 gravitational radii, while the low ionization ultra fast outflow forms a more distant arc beyond about 4000 gravitational radii. }
  }\label{fig:UFO_geometry}
\end{figure}

\citet{Firstpaper2025} estimated that the high-ionization UFO in PDS~456 is located at a distance of $\sim 200\text{--}600\,R_\mathrm{g}$ from the SMBH. Our result of $r \gtrsim \num{4e3}\,R_\mathrm{g}$ for the low-ionization UFO therefore suggests that it is located farther from the SMBH than the high-ionization UFO. 
\revistwomaj{The low-ionization UFO can be shielded by the inner high-ionization UFO, and the \textsc{pion} analysis of \citet{Xu2025} indicates that such screening can modify the SED incident on the soft-X-ray UFO. Nevertheless, for the following order-of-magnitude comparison, we assume that the resulting change in the ionizing luminosity does not substantially affect the comparison and adopt comparable ionizing luminosities for the two phases.} \revistwomaj{Using $\xi \equiv L_\mathrm{ion}/(nr^2)$, together with $\xi = 10^{4.9}\,\unit{\erg\cm\per\s}$ for the high-ionization UFO, $\xi = 10^{3.1}\,\unit{\erg\cm\per\s}$ for the low-ionization UFO, and $r_\mathrm{low}\gtrsim\num{4e3}\,R_\mathrm{g}$, \revis{we find that $n_\mathrm{low}/n_\mathrm{high}$ is at most of order unity.}} \citet{Firstpaper2025} estimated a clump size of $d_\mathrm{clump}=(1.5$--$12)\times10^{14}\,\mathrm{cm}$ and a column density of $N_\mathrm{H}=\qty{e23}{\cm^{-2}}$ per clump for the high-ionization UFO. The density of the high-ionization UFO can therefore be estimated as $n_\mathrm{high} \sim N_\mathrm{H}/d_\mathrm{clump} \sim\qty{e8}{\cm^{-3}}$. \revistwomaj{Thus, the low-ionization UFO has $n_\mathrm{low}\lesssim\qty{e8}{\cm^{-3}}$, with comparable densities obtained only if it lies close to its lower-limit distance. Because the column density of the low-ionization UFO, $N_\mathrm{H}\sim\qty{8e22}{\cm^{-2}}$, is also comparable to that of a high-ionization clump, the characteristic line-of-sight size inferred from $d_\mathrm{clump}\sim N_\mathrm{H}/n$ is comparable to the high-ionization clumps, of order $10^{14}$--$10^{15}\,\mathrm{cm}$, in this limiting case, and would be larger for a larger actual distance.} The outflow velocity of the low-ionization UFO ($\sim 0.3c$) is also comparable to that of the high-ionization UFO ($\sim 0.2$--$0.3c$). Figure~\ref{fig:UFO_geometry} illustrates the resulting structural picture: the high- and low-ionization UFOs have comparable outflow velocities but different ionization states and distances from the SMBH; their number densities and clump sizes can also be comparable if the low-ionization UFO lies near its lower-limit distance. \revistwomaj{This comparison suggests a possible picture in which clumps with similar physical properties propagate outward and are observed at different radii. In this picture, the clumps can retain comparable column densities, number densities, sizes, and velocities near the minimum allowed low-UFO distance, while their ionization state changes naturally with distance through $\xi=L_\mathrm{ion}/(nr^2)$. If the global number of clumps participating in the flow is also comparable, the mass outflow rates of the two phases should be comparable to order of magnitude. Because their outflow velocities are similar, their kinetic energy outflow rates, $\dot{E}_\mathrm{K}\sim \dot{M}_\mathrm{out}v_\mathrm{out}^2/2$, would also be comparable.}

\subsection{Constraining the launching mechanism}
Based on the comparison in Section~\ref{subsubsec:comparison_high_low}, the high- and low-ionization UFOs in PDS~456 have comparable outflow velocities ($\sim 0.2$--$0.3c$) but are inferred to reside at different distances from the SMBH. This observed velocity--distance structure allows us to discuss the launching mechanism. In the radiation-pressure-driven scenario, the outflow velocity profile along a streamline obeys the CAK law \citep {CAK_1975}:
\begin{equation}
  v_\mathrm{out}(r) \propto \left(1 - \frac{R_0}{r}\right)^\beta,
\end{equation}
where $R_0$ is the launching radius, and $\beta$ is a parameter that determines the acceleration profile. For typical values of $\beta \sim 0.5$--$1$\revistwo{\citep[e.g.,][]{CAK_1975,Hagino2015}}, the outflow velocity increases with distance from the launching point and approaches $v_\infty$ at large distances.
In the magnetocentrifugally driven scenario, one may instead consider how the characteristic outflow velocity scales with launching radius. In self-similar MHD wind models, where the large-scale magnetic field and streamline geometry are assumed to be self-similar over the disk radii \citep {Fukumura2010}, the characteristic outflow velocity associated with a streamline launched from radius $r$ can be written as
\begin{equation}
  v_\mathrm{out}(r) \propto r^{-1/2}.
\end{equation}
This reflects the Keplerian velocity profile of the disk, $v_\mathrm{K} \propto r^{-1/2}$: because the outflow is tied to the disk through magnetic field lines and extracts angular momentum from it, the terminal velocity is expected to be proportional to the Keplerian velocity at the launching radius. In this sense, unlike the CAK law, this relation is more naturally interpreted as a comparison among streamlines launched from different radii, with larger launch radii corresponding to slower outflows.\par
\begin{figure}[tbp]
  \begin{center}
    \includegraphics[width=8cm]{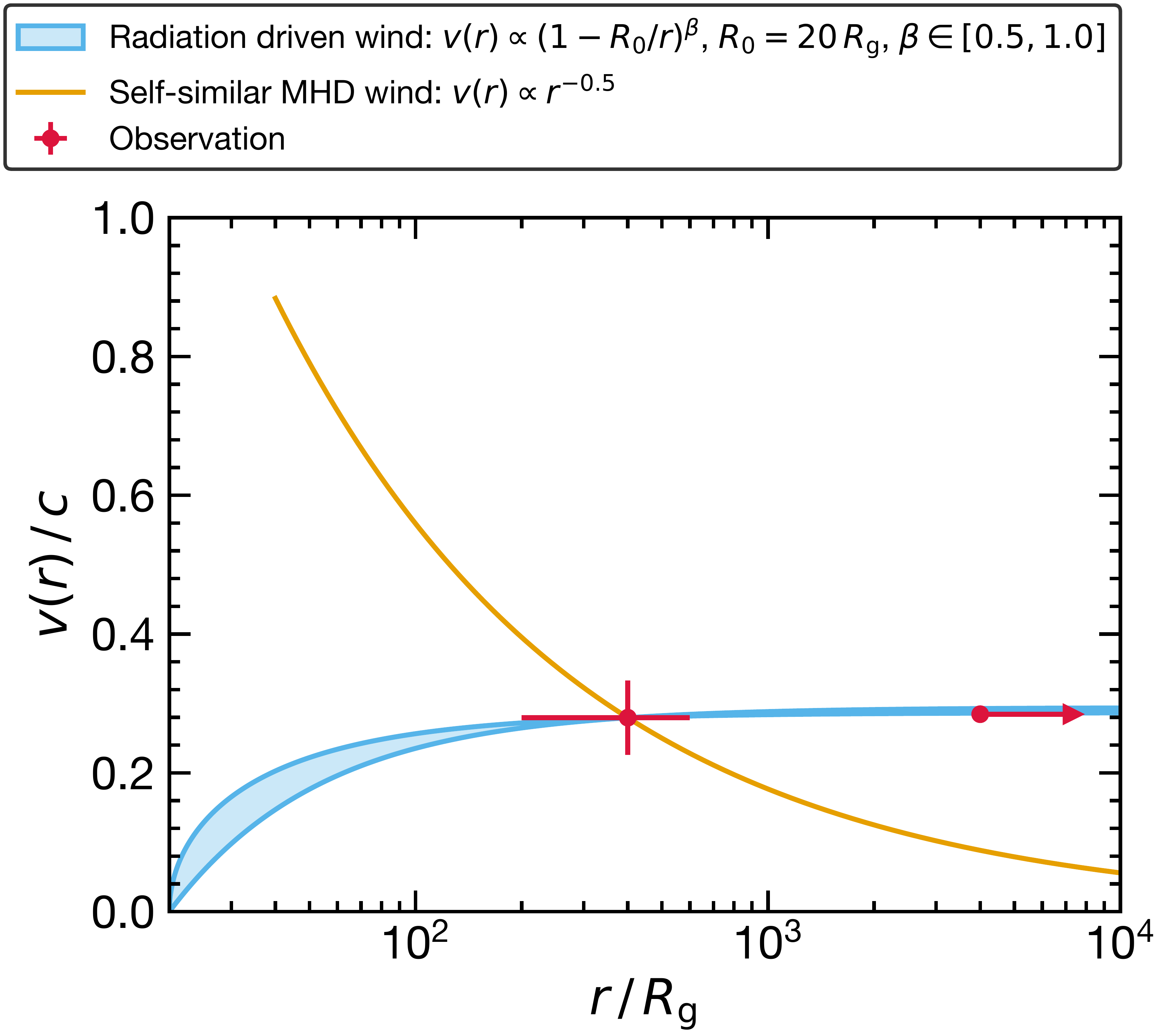}
  \end{center}
  \caption{\revisone{Comparison of the observed velocity--distance relation of the high- and low-ionization UFOs with driving-mechanism models. For the observational data points, the red leftward arrow marks the lower limit on the low-ionization UFO distance, while the horizontal error bar for the high-ionization UFO reflects the velocity spread of the five clumps. }
  {Alt text: Line graph with distance from about 20 to 10000 gravitational radii on the logarithmic x axis and velocity from zero to one times the speed of light on the y axis. The two observational constraints lie near 0.3 times the speed of light, while the model curves show different radial trends. }
  }\label{fig:launching_mechanism}
\end{figure}

\revistwo{Figure~\ref{fig:launching_mechanism} shows the observed velocity--distance relation inferred from the high- and low-ionization UFO components, together with the CAK law and the self-similar MHD wind model.}
The fact that these two UFO phases show similar outflow velocities despite being located at different distances is naturally consistent with the CAK law for $\beta \sim 0.5$--1. Therefore, the observed velocity--distance structure of the UFOs in PDS~456 favors the radiation-pressure-driven scenario. \revistwomaj{For UV line driving to accelerate the wind efficiently, additional lower-ionization material with substantial UV line opacity is required closer to the SMBH, where it must be protected from X-ray overionization by shielding gas. Although absorption from such additional lower-ionization material is not detected in the present observation, this does not conflict with the line-driven interpretation because its visibility depends on the viewing angle. \citet{Mizumoto2021} showed that some sightlines through the upper, highly ionized part of a UV line-driven disk wind do not intercept the lower-ionization acceleration region located closer to the disk.} In contrast, the observed velocity--distance structure is inconsistent with a self-similar magnetocentrifugal wind in which the high- and low-ionization UFOs are launched from a wide range of disk radii spanning $\sim 10^2$--$10^3\,R_\mathrm{g}$ and retain the relation $v \propto r^{-1/2}$. \par
\revistwo{We note that} more general MHD-wind interpretations are still possible. One such possibility is that the high- and low-ionization UFOs trace different locations along the same streamline, rather than streamlines launched from different disk radii. This situation may arise if a substantial part of the magnetic energy is dissipated over a compact region, as in magnetic reconnection \revistwo{\citep{Gu2025a, Reeves2026, Condo2026a}}, so that magnetic acceleration has already saturated by the high-ionization UFO radius. The subsequent flow is then approximately ballistic, and conservation of specific mechanical energy gives
\begin{equation}
  \frac{1}{2}\left(v_\mathrm{high}^2-v_\mathrm{low}^2\right)
  =
  GM_\mathrm{BH}\left(\frac{1}{r_\mathrm{high}}-\frac{1}{r_\mathrm{low}}\right)
\end{equation}
where $v_\mathrm{high}$ and $v_\mathrm{low}$ are the outflow velocities of the high- and low-ionization UFOs, and $r_\mathrm{high}$ and $r_\mathrm{low}$ are their respective distances from the SMBH.
For $r_\mathrm{high}=\num{200}$--$\num{600}\,R_\mathrm{g}$ and $r_\mathrm{low}\geq\num{4000}\,R_\mathrm{g}$, this relation gives $v_\mathrm{high}\simeq 0.30$--$0.31c$ for $v_\mathrm{low}=0.298c$. Once the flow has become inertia dominated, the high- and low-ionization UFOs can therefore show nearly the same velocity along a single MHD streamline.
\par
The next observational step is to apply the distance-estimation method developed here from the crossing velocity of the low-ionization UFO to each of the multiple velocity clumps of the high-ionization UFO. Since these high-ionization clumps are already resolved by \xrism/\resolve\ but do not show significant variability in the present observation, longer-duration \xrism/\resolve\ monitoring would be the most direct way to search for their covering-fraction variability. \revisone{Future} missions with a much larger effective area such as \textit{NewAthena} \revis{\citep{NewAthenaConcept2025}} would further improve the time-resolved measurements. Such observations would provide the transverse velocity of each high-ionization clump, in addition to the line-of-sight velocity, $v_\mathrm{los}$, already measured from the absorption-line shifts.
The key new information would be the relative ordering of the clumps with distance from the SMBH. \revisone{A related ordering problem was explored by \citet{Xu2025}, who compared possible absorber sequences in photoionization modeling to account for screening among multiple layers. The transverse-velocity measurements proposed here would provide an independent geometrical probe of the same ordering.}
If the transverse motion mainly traces the rotational motion of the absorber, then in an inertia-dominated region with weak magnetic torque, a smaller $v_\mathrm{cross}$ would indicate a larger distance from the SMBH. One could then examine whether $v_\mathrm{los}$ increases or decreases along this distance sequence. An outward increase in $v_\mathrm{los}$ would support continued radiation-pressure acceleration, whereas an outward decrease in $v_\mathrm{los}$ would favor inertia-dominated motion after compact magnetic acceleration by magnetic reconnection.


\section{Conclusions}
We analyzed the simultaneous March 2024 observations of PDS~456 obtained with \xrism\ and \nustar\ to investigate the soft X-ray variability associated with the low-ionization UFO. A model-independent comparison between the flare and quiescent phases revealed spectral variability around 1~keV in the rest frame, while the hard X-ray spectral shape remained nearly unchanged. Broadband spectral fitting confirmed that this soft X-ray structure is well described by a partial-covering low-ionization UFO with $\log (\xi/(\unit{\erg\cm\per\s})) \simeq 3.1$ and $v_\mathrm{out}\simeq 0.30c$.\par
The time-sliced spectra showed significant variability in the covering fraction of this absorber. By interpreting this variability as transverse motion across the X-ray source, we converted the soft X-ray spectral variability into a geometrical constraint on the UFO. The resulting crossing velocity is $v_\mathrm{cross}\lesssim \num{5e-3}c$, and combining this limit with angular-momentum conservation and the escape-radius constraint on the launching radius gives a lower limit on the absorber distance of $r \gtrsim \num{4e3}\,R_\mathrm{g}$.\par
This distance is substantially larger than the $\sim 200$--$600\,R_\mathrm{g}$ location inferred for the high-ionization UFO, although the two phases have comparable outflow velocities. Combining the distance constraint obtained here from the transverse motion of the low-ionization UFO with the previously inferred distance of the high-ionization UFO, together with the absorption-line velocities of both phases, places the two UFO phases on the velocity--distance plane. \revistwo{The resulting two-phase structure disfavors a self-similar magnetocentrifugal wind extending over these radii, but is consistent with either radiation-pressure acceleration following a CAK-like velocity law or compact magnetic acceleration through magnetic reconnection followed by nearly ballistic propagation.}




\begin{ack}
\revis{We thank the anonymous referee for a careful review and useful comments that improved this paper.}
This work was supported by the International Graduate Program for Excellence in Earth-Space Science (IGPEES), a World-leading Innovative Graduate Study (WINGS) Program, the University of Tokyo (RS), and by JSPS KAKENHI Grant Number JP26KJ0942 (RS).
\revis{This research was supported by a grant from the Hayakawa Satio Fund awarded by the Astronomical Society of Japan (RS) and funding from the Japan Foundation for Promotion of Astronomy (RS).}
Additional support was provided by JSPS KAKENHI Grant Numbers JP21K13963 and JP24K00638 (KH), JP21K13958 and JP26K17197 (MM), and JP23K25907 (AB).
This work was also supported by the Yamada Science Foundation (MM), NASA under Award Number 80GSFC24M0006 (TY), the Israel Science Foundation under Grant Number 2617/25 (EB), and the Canadian Space Agency under Grant Number 25EXPRSM1 (LCG).
\end{ack}



\bibliographystyle{apj}
\bibliography{PDS456_covfactor_variation_PASJ}

@Article{Nardini2015,
  author    = {Nardini, E. and Reeves, J. N. and Gofford, J. and Harrison, F. A. and Risaliti, G. and Braito, V. and Costa, M. T. and Matzeu, G. A. and Walton, D. J. and Behar, E. and Boggs, S. E. and Christensen, F. E. and Craig, W. W. and Hailey, C. J. and Matt, G. and Miller, J. M. and O’Brien, P. T. and Stern, D. and Turner, T. J. and Ward, M. J.},
  journal   = {Science},
  title     = {Black hole feedback in the luminous quasar PDS 456},
  year      = {2015},
  issn      = {1095-9203},
  month     = feb,
  number    = {6224},
  pages     = {860--863},
  volume    = {347},
  doi       = {10.1126/science.1259202},
  publisher = {American Association for the Advancement of Science (AAAS)},
}

@ARTICLE{Tombesi2012,
       author = {{Tombesi}, F. and {Cappi}, M. and {Reeves}, J.~N. and {Braito}, V.},
        title = "{Evidence for ultrafast outflows in radio-quiet AGNs - III. Location and energetics}",
      journal = {MNRAS},
         year = 2012,
        month = may,
       volume = {422},
       number = {1},
        pages = {L1-L5},
          doi = {10.1111/j.1745-3933.2012.01221.x},
archivePrefix = {arXiv},
       eprint = {1201.1897},
 primaryClass = {astro-ph.HE},
}

@ARTICLE{Hagino2015,
       author = {{Hagino}, Kouichi and {Odaka}, Hirokazu and {Done}, Chris and {Gandhi}, Poshak and {Watanabe}, Shin and {Sako}, Masao and {Takahashi}, Tadayuki},
        title = "{The origin of ultrafast outflows in AGN: Monte Carlo simulations of the wind in PDS 456}",
      journal = {MNRAS},
         year = 2015,
        month = jan,
       volume = {446},
       number = {1},
        pages = {663-676},
          doi = {10.1093/mnras/stu2095},
archivePrefix = {arXiv},
       eprint = {1410.1640},
 primaryClass = {astro-ph.HE},
}

@ARTICLE{Fukumura2015,
       author = {{Fukumura}, Keigo and {Tombesi}, Francesco and {Kazanas}, Demosthenes and {Shrader}, Chris and {Behar}, Ehud and {Contopoulos}, Ioannis},
        title = "{Magnetically Driven Accretion Disk Winds and Ultra-fast Outflows in PG 1211+143}",
      journal = {ApJ},
         year = 2015,
        month = may,
       volume = {805},
       number = {1},
          eid = {17},
        pages = {17},
          doi = {10.1088/0004-637X/805/1/17},
archivePrefix = {arXiv},
       eprint = {1503.04074},
 primaryClass = {astro-ph.HE},
}

@article{Firstpaper2025,
  title = {Structured Ionized Winds Shooting out from a Quasar at Relativistic Speeds},
  author = {{XRISM collaboration} and Audard, Marc and Awaki, Hisamitsu and Ballhausen, Ralf and Bamba, Aya and Behar, Ehud and {Boissay-Malaquin}, Rozenn and Brenneman, Laura and Brown, Gregory V. and Corrales, Lia and Costantini, Elisa and Cumbee, Renata and Trigo, Mar{\'i}a D{\'i}az and Done, Chris and Dotani, Tadayasu and Ebisawa, Ken and Eckart, Megan and Eckert, Dominique and Enoto, Teruaki and Eguchi, Satoshi and Ezoe, Yuichiro and Foster, Adam and Fujimoto, Ryuichi and Fujita, Yutaka and Fukazawa, Yasushi and Fukushima, Kotaro and Furuzawa, Akihiro and Gallo, Luigi and Garc{\'i}a, Javier A. and Gu, Liyi and Guainazzi, Matteo and Hagino, Kouichi and Hamaguchi, Kenji and Hatsukade, Isamu and Hayashi, Katsuhiro and Hayashi, Takayuki and Hell, Natalie and {Hodges-Kluck}, Edmund and Hornschemeier, Ann and Ichinohe, Yuto and Ishida, Manabu and Ishikawa, Kumi and Ishisaki, Yoshitaka and Kaastra, Jelle and Kallman, Timothy and Kara, Erin and Katsuda, Satoru and Kanemaru, Yoshiaki and Kelley, Richard and Kilbourne, Caroline and Kitamoto, Shunji and Kobayashi, Shogo and Kohmura, Takayoshi and Kubota, Aya and Leutenegger, Maurice and Loewenstein, Michael and Maeda, Yoshitomo and Markevitch, Maxim and Matsumoto, Hironori and Matsushita, Kyoko and McCammon, Dan and McNamara, Brian and Mernier, Fran{\c c}ois and Miller, Eric D. and Miller, Jon M. and Mitsuishi, Ikuyuki and Mizumoto, Misaki and Mizuno, Tsunefumi and Mori, Koji and Mukai, Koji and Murakami, Hiroshi and Mushotzky, Richard and Nakajima, Hiroshi and Nakazawa, Kazuhiro and Ness, Jan-Uwe and Nobukawa, Kumiko and Nobukawa, Masayoshi and Noda, Hirofumi and Odaka, Hirokazu and Ogawa, Shoji and Ogorzalek, Anna and Okajima, Takashi and Ota, Naomi and Paltani, Stephane and Petre, Robert and Plucinsky, Paul and Porter, Frederick Scott and Pottschmidt, Katja and Sato, Kosuke and Sato, Toshiki and Sawada, Makoto and Seta, Hiromi and Shidatsu, Megumi and Simionescu, Aurora and Smith, Randall and Suzuki, Hiromasa and Szymkowiak, Andrew and Takahashi, Hiromitsu and Takeo, Mai and Tamagawa, Toru and Tamura, Keisuke and Tanaka, Takaaki and Tanimoto, Atsushi and Tashiro, Makoto and Terada, Yukikatsu and Terashima, Yuichi and Tsuboi, Yohko and Tsujimoto, Masahiro and Tsunemi, Hiroshi and Tsuru, Takeshi G. and Uchida, Hiroyuki and Uchida, Nagomi and Uchida, Yuusuke and Uchiyama, Hideki and Ueda, Yoshihiro and Uno, Shinichiro and Vink, Jacco and Watanabe, Shin and Williams, Brian J. and Yamada, Satoshi and Yamada, Shinya and Yamaguchi, Hiroya and Yamaoka, Kazutaka and Yamasaki, Noriko and Yamauchi, Makoto and Yamauchi, Shigeo and Yaqoob, Tahir and Yoneyama, Tomokage and Yoshida, Tessei and Yukita, Mihoko and Zhuravleva, Irina and Braito, Valentina and Cond{\`o}, Pierpaolo and Fukumura, Keigo and Gonzalez, Adam and Luminari, Alfredo and Miyamoto, Aiko and Mizukawa, Ryuki and Reeves, James and Sato, Riki and Tombesi, Francesco and Xu, Yerong},
  year = 2025,
  month = may,
  journal = {Nature},
  issn = {0028-0836, 1476-4687},
  doi = {10.1038/s41586-025-08968-2},
  urldate = {2025-05-16},
  langid = {english},
}

@article{Amorim2024,
  title = {The Size-Luminosity Relation of Local Active Galactic Nuclei from Interferometric Observations of the Broad-Line Region},
  author = {Amorim, A. and Bourdarot, G. and Brandner, W. and Cao, Y. and Cl{\'e}net, Y. and Davies, R. and de Zeeuw, P. T. and Dexter, J. and Drescher, A. and Eckart, A. and Eisenhauer, F. and Fabricius, M. and Feuchtgruber, H. and Schreiber, N. M. F{\"o}rster and Garcia, P. J. V. and Genzel, R. and Gillessen, S. and Gratadour, D. and H{\"o}nig, S. and Kishimoto, M. and Lacour, S. and Lutz, D. and Millour, F. and Netzer, H. and Ott, T. and Paumard, T. and Perraut, K. and Perrin, G. and Peterson, B. M. and Petrucci, P. O. and Pfuhl, O. and Prieto, M. A. and Rabien, S. and Rouan, D. and Santos, D. J. D. and Shangguan, J. and Shimizu, T. and Sternberg, A. and Straubmeier, C. and Sturm, E. and Tacconi, L. J. and Tristram, K. R. W. and Widmann, F. and Woillez, J.},
  year = 2024,
  month = apr,
  journal = {A\&A},
  volume = {684},
  pages = {A167},
  publisher = {EDP Sciences},
  issn = {0004-6361, 1432-0746},
  doi = {10.1051/0004-6361/202348167},
  urldate = {2025-07-27},
  copyright = {\copyright{} The Authors 2024},
  langid = {english},
}

@article{Reeves2020_lowUFO,
  title = {Resolving the {{Soft X-Ray Ultrafast Outflow}} in {{PDS}} 456},
  author = {Reeves, J. N. and Braito, V. and Chartas, G. and Hamann, F. and Laha, S. and Nardini, E.},
  year = 2020,
  month = may,
  journal = {ApJ},
  volume = {895},
  number = {1},
  pages = {37},
  publisher = {The American Astronomical Society},
  issn = {0004-637X},
  doi = {10.3847/1538-4357/ab8cc4},
  langid = {english},
}

@article{CAK_1975,
  title = {Radiation-Driven Winds in {{Of}} Stars},
  author = {Castor, J. I. and Abbott, D. C. and Klein, R. I.},
  year = 1975,
  month = jan,
  journal = {ApJ},
  volume = {195},
  pages = {157},
  issn = {0004-637X, 1538-4357},
  doi = {10.1086/153315},
  langid = {english},
}

@article{wilms_abund,
  title = {On the {{Absorption}} of {{X-Rays}} in {{theInterstellar Medium}}},
  author = {Wilms, J. and Allen, A. and McCray, R.},
  year = 2000,
  month = oct,
  journal = {ApJ},
  volume = {542},
  number = {2},
  pages = {914},
  publisher = {IOP Publishing},
  issn = {0004-637X},
  doi = {10.1086/317016},
  langid = {english},
}

@article{gofford_2014_lowUFO,
  title = {{{REVEALING THE LOCATION AND STRUCTURE OF THE ACCRETION DISK WIND IN PDS}} 456},
  author = {Gofford, J. and Reeves, J. N. and Braito, V. and Nardini, E. and Costa, M. T. and Matzeu, G. A. and O'Brien, P. and Ward, M. and Turner, T. J. and Miller, L.},
  year = 2014,
  month = mar,
  journal = {ApJ},
  volume = {784},
  number = {1},
  pages = {77},
  issn = {0004-637X, 1538-4357},
  doi = {10.1088/0004-637X/784/1/77},
  langid = {english},
}

@article{reeves_2016_low_UFO,
  title = {{{DISCOVERY OF BROAD SOFT X-RAY ABSORPTION LINES FROM THE QUASAR WIND IN PDS}} 456},
  author = {Reeves, J. N. and Braito, V. and Nardini, E. and Behar, E. and O'Brien, P. T. and Tombesi, F. and Turner, T. J. and Costa, M. T.},
  year = 2016,
  month = jun,
  journal = {ApJ},
  volume = {824},
  number = {1},
  pages = {20},
  publisher = {The American Astronomical Society},
  issn = {0004-637X},
  doi = {10.3847/0004-637X/824/1/20},
  langid = {english},
}

@article{Magorrian1998,
  title = {The {{Demography}} of {{Massive Dark Objects}} in {{GalaxyCenters}}},
  author = {Magorrian, John and Tremaine, Scott and Richstone, Douglas and Bender, Ralf and Bower, Gary and Dressler, Alan and Faber, S. M. and Gebhardt, Karl and Green, Richard and Grillmair, Carl and Kormendy, John and Lauer, Tod},
  year = 1998,
  month = jun,
  journal = {AJ},
  volume = {115},
  number = {6},
  pages = {2285},
  publisher = {IOP Publishing},
  issn = {1538-3881},
  doi = {10.1086/300353},
  urldate = {2026-05-07},
  langid = {english},
}

@article{Ferrarese2000,
  title = {A {{Fundamental Relation}} between {{Supermassive Black Holes}} and {{TheirHost Galaxies}}},
  author = {Ferrarese, Laura and Merritt, David},
  year = 2000,
  month = aug,
  journal = {ApJ},
  volume = {539},
  number = {1},
  pages = {L9},
  publisher = {IOP Publishing},
  issn = {0004-637X},
  doi = {10.1086/312838},
  urldate = {2026-05-07},
  langid = {english},
}

@article{Kormendy2013,
  title = {Coevolution ({{Or Not}}) of {{Supermassive Black Holes}} and {{Host Galaxies}}},
  author = {Kormendy, John and Ho, Luis C.},
  year = 2013,
  month = aug,
  journal = {ARA\&A},
  volume = {51},
  number = {Volume 51, 2013},
  pages = {511--653},
  publisher = {Annual Reviews},
  issn = {0066-4146, 1545-4282},
  doi = {10.1146/annurev-astro-082708-101811},
  urldate = {2026-05-08},
  langid = {english},
}

@ARTICLE{Silk1998,
       author = {{Silk}, Joseph and {Rees}, Martin J.},
        title = "{Quasars and galaxy formation}",
      journal = {A\&A},
         year = 1998,
        month = mar,
       volume = {331},
        pages = {L1-L4},
          doi = {10.48550/arXiv.astro-ph/9801013},
archivePrefix = {arXiv},
       eprint = {astro-ph/9801013},
 primaryClass = {astro-ph},
}

@article{King2003,
  title = {Black {{Holes}}, {{Galaxy Formation}}, and the {{MBH-$\sigma$ Relation}}},
  author = {King, Andrew},
  year = 2003,
  month = sep,
  journal = {ApJ},
  volume = {596},
  number = {1},
  pages = {L27},
  publisher = {IOP Publishing},
  issn = {0004-637X},
  doi = {10.1086/379143},
  urldate = {2026-05-08},
  langid = {english},
}

@article{Fabian2012a,
  title = {Observational {{Evidence}} of {{Active Galactic Nuclei Feedback}}},
  author = {Fabian, A. C.},
  year = 2012,
  month = sep,
  journal = {ARA\&A},
  volume = {50},
  number = {Volume 50, 2012},
  pages = {455--489},
  publisher = {Annual Reviews},
  issn = {0066-4146, 1545-4282},
  doi = {10.1146/annurev-astro-081811-125521},
  urldate = {2026-05-08},
  langid = {english},
}

@article{Pounds2003,
  title = {A High-Velocity Ionized Outflow and {{XUV}} Photosphere in the Narrow Emission Line Quasar {{PG1211}}+143},
  author = {Pounds, K. A. and Reeves, J. N. and King, A. R. and Page, K. L. and O'Brien, P. T. and Turner, M. J. L.},
  year = 2003,
  month = nov,
  journal = {MNRAS},
  volume = {345},
  number = {3},
  pages = {705--713},
  issn = {00358711, 13652966},
  doi = {10.1046/j.1365-8711.2003.07006.x},
  urldate = {2025-12-30},
  langid = {english},
}

@article{Tombesi2010,
  title = {Evidence for Ultra-Fast Outflows in Radio-Quiet {{AGNs}}: {{I}} - Detection and Statistical Incidence of {{Fe K-shell}} Absorption Lines},
  shorttitle = {Evidence for Ultra-Fast Outflows in Radio-Quiet {{AGNs}}},
  author = {Tombesi, F. and Cappi, M. and Reeves, J. N. and Palumbo, G. G. C. and Yaqoob, T. and Braito, V. and Dadina, M.},
  year = 2010,
  month = oct,
  journal = {A\&A},
  volume = {521},
  eprint = {1006.2858},
  primaryclass = {astro-ph},
  pages = {A57},
  issn = {0004-6361, 1432-0746},
  doi = {10.1051/0004-6361/200913440},
  urldate = {2025-04-16},
}

@article{Gofford2013,
  title = {The {{Suzaku}} View of Highly Ionized Outflows in {{AGN}} -- {{I}}. {{Statistical}} Detection and Global Absorber Properties},
  author = {Gofford, Jason and Reeves, James N. and Tombesi, Francesco and Braito, Valentina and Turner, T. Jane and Miller, Lance and Cappi, Massimo},
  year = 2013,
  month = mar,
  journal = {MNRAS},
  volume = {430},
  number = {1},
  pages = {60--80},
  issn = {0035-8711},
  doi = {10.1093/mnras/sts481},
  urldate = {2025-05-02},
  langid = {american},
}

@article{KingPounds_UFO2003,
  title = {Black Hole Winds},
  author = {King, A. R. and Pounds, K. A.},
  year = 2003,
  month = oct,
  journal = {MNRAS},
  volume = {345},
  number = {2},
  pages = {657--659},
  issn = {0035-8711, 1365-2966},
  doi = {10.1046/j.1365-8711.2003.06980.x},
  urldate = {2026-05-08},
  langid = {english},
}

@article{Takeuchi2013,
  title = {Clumpy {{Outflows}} from {{Supercritical Accretion Flow}}},
  author = {Takeuchi, Shun and Ohsuga, Ken and Mineshige, Shin},
  year = 2013,
  month = aug,
  journal = {PASJ},
  volume = {65},
  number = {4},
  pages = {88},
  issn = {2053-051X, 0004-6264},
  doi = {10.1093/pasj/65.4.88},
  urldate = {2026-05-08},
  langid = {english},
}

@article{Nomura2016,
  title = {Radiation Hydrodynamic Simulations of Line-Driven Disk Winds for Ultra-Fast Outflows},
  author = {Nomura, Mariko and Ohsuga, Ken and Takahashi, Hiroyuki R. and Wada, Keiichi and Yoshida, Tessei},
  year = 2016,
  month = feb,
  journal = {PASJ},
  volume = {68},
  number = {1},
  pages = {16},
  issn = {2053-051X, 0004-6264},
  doi = {10.1093/pasj/psv124},
  urldate = {2026-05-08},
  langid = {english},
}

@article{Mizumoto2021,
  title = {{{UV}} Line Driven Disc Wind as the Origin of Ultrafast Outflows in {{AGN}}},
  author = {Mizumoto, Misaki and Nomura, Mariko and Done, Chris and Ohsuga, Ken and Odaka, Hirokazu},
  year = 2021,
  month = mar,
  journal = {MNRAS},
  volume = {503},
  number = {1},
  eprint = {2003.01137},
  primaryclass = {astro-ph},
  pages = {1442--1458},
  issn = {0035-8711, 1365-2966},
  doi = {10.1093/mnras/staa3282},
  urldate = {2025-03-11},
  archiveprefix = {arXiv},
  langid = {american},
}

@article{Blandford1982,
  title = {Hydromagnetic Flows from Accretion Discs and the Production of Radio Jets},
  author = {Blandford, R. D. and Payne, D. G.},
  year = 1982,
  month = aug,
  journal = {MNRAS},
  volume = {199},
  number = {4},
  pages = {883--903},
  issn = {0035-8711, 1365-2966},
  doi = {10.1093/mnras/199.4.883},
  urldate = {2026-05-08},
  langid = {english},
}

@article{Contopoulos1994,
  title = {Magnetically Driven Jets and Winds: {{Exact}} Solutions},
  shorttitle = {Magnetically Driven Jets and Winds},
  author = {Contopoulos, J. and Lovelace, R. V. E.},
  year = 1994,
  month = jul,
  journal = {ApJ},
  volume = {429},
  pages = {139},
  issn = {0004-637X, 1538-4357},
  doi = {10.1086/174307},
  urldate = {2026-05-08},
  langid = {english},
}

@article{Fukumura2010,
  title = {{{MODELING HIGH-VELOCITY QSO ABSORBERS WITH PHOTOIONIZED MAGNETOHYDRODYNAMIC DISK WINDS}}},
  author = {Fukumura, Keigo and Kazanas, Demosthenes and Contopoulos, Ioannis and Behar, Ehud},
  year = 2010,
  month = oct,
  journal = {ApJL},
  volume = {723},
  number = {2},
  pages = {L228},
  publisher = {The American Astronomical Society},
  issn = {2041-8205},
  doi = {10.1088/2041-8205/723/2/L228},
  urldate = {2025-05-03},
  langid = {english},
}

@article{Gu2025a,
  title = {Delving into the Depths of {{NGC}} 3783 with {{XRISM}}: {{III}}. {{Birth}} of an Ultrafast Outflow during a Soft Flare},
  shorttitle = {Delving into the Depths of {{NGC}} 3783 with {{XRISM}}},
  author = {Gu, Liyi and Fukumura, Keigo and Kaastra, Jelle and Eckart, Megan and Ballhausen, Ralf and Behar, Ehud and Diez, Camille and Guainazzi, Matteo and Kallman, Timothy and Kara, Erin and Li, Chen and Mehdipour, Missagh and Mizumoto, Misaki and Ogawa, Shoji and Panagiotou, Christos and Signorini, Matilde and Tanimoto, Atsushi and Zhao, Keqin and Noda, Hirofumi and Miller, Jon and Yamada, Satoshi},
  year = 2025,
  month = dec,
  journal = {A\&A},
  volume = {704},
  pages = {A146},
  issn = {0004-6361, 1432-0746},
  doi = {10.1051/0004-6361/202557189},
  urldate = {2026-05-08},
  langid = {english},
}

@article{Reeves2003,
  title = {A {{Massive X-Ray Outflow}} from the {{Quasar PDS}} 456},
  author = {Reeves, J. N. and O'Brien, P. T. and Ward, M. J.},
  year = 2003,
  month = jul,
  journal = {ApJ},
  volume = {593},
  number = {2},
  pages = {L65},
  publisher = {IOP Publishing},
  issn = {0004-637X},
  doi = {10.1086/378218},
  urldate = {2026-05-08},
  langid = {english},
}

@article{Matzeu2016,
  title = {Short-Term {{X-ray}} Spectral Variability of the Quasar {{PDS}}~456 Observed in a Low-Flux State},
  author = {Matzeu, G. A. and Reeves, J. N. and Nardini, E. and Braito, V. and Costa, M. T. and Tombesi, F. and Gofford, J.},
  year = 2016,
  month = may,
  journal = {MNRAS},
  volume = {458},
  number = {2},
  pages = {1311--1329},
  issn = {0035-8711},
  doi = {10.1093/mnras/stw354},
  urldate = {2025-03-14},
  langid = {american},
}

@article{GravityCollab_2023,
  title = {Toward Measuring Supermassive Black Hole Masses with Interferometric Observations of the Dust Continuum},
  author = {{GRAVITY Collaboration} and Amorim, A. and Bourdarot, G. and Brandner, W. and Cao, Y. and Cl{\'e}net, Y. and Davies, R. and De Zeeuw, P. T. and Dexter, J. and Drescher, A. and Eckart, A. and Eisenhauer, F. and Fabricius, M. and F{\"o}rster Schreiber, N. M. and Garcia, P. J. V. and Genzel, R. and Gillessen, S. and Gratadour, D. and H{\"o}nig, S. and Kishimoto, M. and Lacour, S. and Lutz, D. and Millour, F. and Netzer, H. and Ott, T. and Paumard, T. and Perraut, K. and Perrin, G. and Peterson, B. M. and Petrucci, P. O. and Pfuhl, O. and Prieto, M. A. and Rouan, D. and Santos, D. J. D. and Shangguan, J. and Shimizu, T. and Sternberg, A. and Straubmeier, C. and Sturm, E. and Tacconi, L. J. and Tristram, K. R. W. and Widmann, F. and Woillez, J.},
  year = 2023,
  month = jan,
  journal = {A\&A},
  volume = {669},
  pages = {A14},
  issn = {0004-6361, 1432-0746},
  doi = {10.1051/0004-6361/202244655},
  urldate = {2026-05-17},
  copyright = {https://creativecommons.org/licenses/by/4.0},
  langid = {english},
}

@article{Midooka2026_lowUFO,
  title = {Indication of the Less-Ionized Clumpy Ultra-Fast Outflows in {{Seyfert}} Galaxies},
  author = {Midooka, Takuya and Mizumoto, Misaki and Ebisawa, Ken},
  year = 2026,
  month = apr,
  journal = {PASJ},
  volume = {78},
  number = {2},
  pages = {583--600},
  issn = {0004-6264, 2053-051X},
  doi = {10.1093/pasj/psag003},
  urldate = {2026-05-18},
  langid = {english},
}

@article{Xu2025,
  title = {Unraveling the Structure of the Stratified Ultra-Fast Outflows in {{PDS}} 456 with {{XRISM}}},
  author = {Xu, Yerong and Gallo, Luigi C and Hagino, Kouichi and Reeves, James N and Tombesi, Francesco and Mizumoto, Misaki and Luminari, Alfredo and Gonzalez, Adam G and Behar, Ehud and {Boissay-Malaquin}, Rozenn and Braito, Valentina and Cond{\'o}, Pierpaolo and Done, Chris and Miyamoto, Aiko and Mizukawa, Ryuki and Odaka, Hirokazu and Sato, Riki and Tanimoto, Atsushi and Tashiro, Makoto and Yaqoob, Tahir and Yamada, Satoshi},
  year = 2025,
  month = sep,
  journal = {PASJ},
  volume = {77},
  number = {Supplement\_1},
  pages = {S223-S241},
  issn = {0004-6264, 2053-051X},
  doi = {10.1093/pasj/psaf070},
  urldate = {2025-10-09},
  copyright = {https://academic.oup.com/journals/pages/open\_access/funder\_policies/chorus/standard\_publication\_model},
  langid = {english},
}

@article{Jin_softexcess_2017,
  title = {Super-{{Eddington QSO RX J0439}}.6-5311 -- {{I}}. {{Origin}} of the Soft {{X-ray}} Excess and Structure of the Inner Accretion Flow},
  author = {Jin, Chichuan and Done, Chris and Ward, Martin},
  year = 2017,
  month = jul,
  journal = {MNRAS},
  volume = {468},
  number = {3},
  pages = {3663--3681},
  issn = {0035-8711, 1365-2966},
  doi = {10.1093/mnras/stx718},
  urldate = {2026-05-25},
  langid = {english},
}

@article{Done_softexcess_2012,
  title = {Intrinsic Disc Emission and the Soft {{X-ray}} Excess in Active Galactic Nuclei: {{The}} Soft {{X-ray}} Excess in {{AGN}}},
  shorttitle = {Intrinsic Disc Emission and the Soft {{X-ray}} Excess in Active Galactic Nuclei},
  author = {Done, Chris and Davis, S. W. and Jin, C. and Blaes, O. and Ward, M.},
  year = 2012,
  month = mar,
  journal = {MNRAS},
  volume = {420},
  number = {3},
  pages = {1848--1860},
  issn = {00358711},
  doi = {10.1111/j.1365-2966.2011.19779.x},
  urldate = {2026-05-25},
  langid = {english},
}

@article{Petrucci_softexcess_2018,
  title = {Testing Warm {{Comptonization}} Models for the Origin of the Soft {{X-ray}} Excess in {{AGNs}}},
  author = {Petrucci, P.-O. and Ursini, F. and De Rosa, A. and Bianchi, S. and Cappi, M. and Matt, G. and Dadina, M. and Malzac, J.},
  year = 2018,
  month = mar,
  journal = {A\&A},
  volume = {611},
  pages = {A59},
  issn = {0004-6361, 1432-0746},
  doi = {10.1051/0004-6361/201731580},
  urldate = {2026-05-25},
  copyright = {https://www.edpsciences.org/en/authors/copyright-and-licensing},
  langid = {english},
}

@article{Serafinelli_2019,
  title = {Multiphase Quasar-Driven Outflows in {{PG}} 1114+445: {{I}}. {{Entrained}} Ultra-Fast Outflows},
  shorttitle = {Multiphase Quasar-Driven Outflows in {{PG}} 1114+445},
  author = {Serafinelli, Roberto and Tombesi, Francesco and Vagnetti, Fausto and Piconcelli, Enrico and Gaspari, Massimo and Saturni, Francesco G.},
  year = 2019,
  month = jul,
  journal = {A\&A},
  volume = {627},
  pages = {A121},
  issn = {0004-6361, 1432-0746},
  doi = {10.1051/0004-6361/201935275},
  urldate = {2026-05-26},
  copyright = {https://www.edpsciences.org/en/authors/copyright-and-licensing},
  langid = {english},
}

@article{Torres_PDS456_red_1997,
  title = {Discovery of a {{Luminous Quasar}} in the {{Nearby Universe}}*},
  author = {Torres, Carlos A. O. and Quast, Germano R. and Coziol, Roger and Jablonski, Francisco and de la Reza, Ramiro and L{\'e}pine, J. R. D. and {Greg{\'o}rio-Hetem}, J.},
  year = 1997,
  month = sep,
  journal = {ApJ},
  volume = {488},
  number = {1},
  pages = {L19},
  publisher = {IOP Publishing},
  issn = {0004-637X},
  doi = {10.1086/310913},
  urldate = {2026-05-26},
  langid = {english},
}

@article{Condo2026a,
  title = {Unveiling the Dynamics of the Ultrafast Outflow in {{IRAS}} 13224-3809 with {{X-ray}} Spectroscopy},
  author = {Cond{\`o}, Pierpaolo and Tombesi, Francesco and Laurenti, Marco and Luminari, Alfredo and Middei, Riccardo and Piconcelli, Enrico and Gaspari, Massimo and Lanzuisi, Giorgio and Serafinelli, Roberto and Tortosa, Alessia and Zappacosta, Luca and Nicastro, Fabrizio},
  year = 2026,
  month = may,
  journal = {A\&A},
  issn = {0004-6361, 1432-0746},
  doi = {10.1051/0004-6361/202659055},
  urldate = {2026-06-03},
  langid = {english},
}

@article{Reeves2026,
  title = {Winds of {{Change}}: {{XRISM Resolve X-Ray Spectroscopy}} of {{NGC}} 4051},
  shorttitle = {Winds of {{Change}}},
  author = {Reeves, James N. and Ogawa, Shoji and Turner, Tracey J. and Braito, Valentina and Yamada, Satoshi and Kraemer, Steven B. and Noda, Hirofumi and Falc{\~a}o, Anna Trindade and Elvis, Martin and Fabbiano, Giuseppina},
  year = 2026,
  month = apr,
  journal = {ApJ},
  volume = {1001},
  number = {2},
  pages = {137},
  publisher = {The American Astronomical Society},
  issn = {0004-637X},
  doi = {10.3847/1538-4357/ae5527},
  urldate = {2026-06-03},
  langid = {english},
}

@article{FukumuraTombesi2019,
  title = {Constraining {{X-Ray Coronal Size}} with {{Transverse Motion}} of {{AGN Ultra-fast Outflows}}},
  author = {Fukumura, Keigo and Tombesi, Francesco},
  year = 2019,
  month = nov,
  journal = {ApJL},
  volume = {885},
  number = {2},
  pages = {L38},
  publisher = {The American Astronomical Society},
  issn = {2041-8205},
  doi = {10.3847/2041-8213/ab5193},
  urldate = {2026-06-09},
  langid = {english},
}

@article{SUBWAYS2023,
  title = {Supermassive {{Black Hole Winds}} in {{X-rays}}: {{SUBWAYS}} - {{I}}. {{Ultra-fast}} Outflows in Quasars beyond the Local {{Universe}}},
  shorttitle = {Supermassive {{Black Hole Winds}} in {{X-rays}}},
  author = {Matzeu, G. A. and Brusa, M. and Lanzuisi, G. and Dadina, M. and Bianchi, S. and Kriss, G. and Mehdipour, M. and Nardini, E. and Chartas, G. and Middei, R. and Piconcelli, E. and Gianolli, V. and Comastri, A. and Longinotti, A. L. and Krongold, Y. and Ricci, F. and Petrucci, P. O. and Tombesi, F. and Luminari, A. and Zappacosta, L. and Miniutti, G. and Gaspari, M. and Behar, E. and Bischetti, M. and Mathur, S. and Perna, M. and Giustini, M. and Grandi, P. and Torresi, E. and Vignali, C. and Bruni, G. and Cappi, M. and Costantini, E. and Cresci, G. and Marco, B. De and Rosa, A. De and Gilli, R. and Guainazzi, M. and Kaastra, J. and Kraemer, S. and Franca, F. La and Marconi, A. and Panessa, F. and Ponti, G. and Proga, D. and Ursini, F. and Baldini, P. and Fiore, F. and King, A. R. and Maiolino, R. and Matt, G. and Merloni, A.},
  year = 2023,
  month = feb,
  journal = {A\&A},
  volume = {670},
  pages = {A182},
  publisher = {EDP Sciences},
  issn = {0004-6361, 1432-0746},
  doi = {10.1051/0004-6361/202245036},
  urldate = {2026-07-24},
  copyright = {\copyright{} The Authors 2023},
  langid = {english},
}

@inproceedings{XSPEC1996a,
  title = {{{XSPEC}}: {{The First Ten Years}}},
  shorttitle = {{{XSPEC}}},
  booktitle = {Astronomical {{Data Analysis Software}} and {{Systems V}}},
  author = {Arnaud, K. A.},
  year = 1996,
  month = jan,
  volume = {101},
  pages = {17},
  urldate = {2026-08-06},
}

@article{XSTAR2001,
  title = {Photoionization and {{High}}-{{Density Gas}}},
  author = {Kallman, T. and Bautista, M.},
  year = 2001,
  month = mar,
  journal = {ApJS},
  volume = {133},
  number = {1},
  pages = {221--253},
  issn = {0067-0049, 1538-4365},
  doi = {10.1086/319184},
  urldate = {2026-08-06},
}

@article{Cash1979,
  title = {Parameter Estimation in Astronomy through Application of the Likelihood Ratio},
  author = {Cash, W.},
  year = 1979,
  month = mar,
  journal = {ApJ},
  volume = {228},
  pages = {939},
  issn = {0004-637X, 1538-4357},
  doi = {10.1086/156922},
  urldate = {2026-08-10},
}

@article{Wilks1938,
  title = {The {{Large-Sample Distribution}} of the {{Likelihood Ratio}} for {{Testing Composite Hypotheses}}},
  author = {Wilks, S. S.},
  year = 1938,
  month = mar,
  journal = {Ann. Math. Stat.},
  volume = {9},
  number = {1},
  pages = {60--62},
  issn = {0003-4851},
  doi = {10.1214/aoms/1177732360},
  urldate = {2026-08-10},
}

@article{NewAthenaConcept2025,
  title = {The {{NewAthena}} Mission Concept in the Context of the next Decade of {{X-ray}} Astronomy},
  author = {Cruise, Mike and Guainazzi, Matteo and Aird, James and Carrera, Francisco J. and Costantini, Elisa and Corrales, Lia and Dauser, Thomas and Eckert, Dominique and Gastaldello, Fabio and Matsumoto, Hironori and Osten, Rachel and Petrucci, Pierre-Olivier and Porquet, Delphine and Pratt, Gabriel W. and Rea, Nanda and Reiprich, Thomas H. and Simionescu, Aurora and Spiga, Daniele and Troja, Eleonora},
  year = 2025,
  month = jan,
  journal = {Nat. Astron.},
  volume = {9},
  number = {1},
  pages = {36--44},
  publisher = {Nature Publishing Group},
  issn = {2397-3366},
  doi = {10.1038/s41550-024-02416-3},
  urldate = {2026-08-10},
  copyright = {2024 Springer Nature Limited},
  langid = {english},
}

@article{KingPounds_Feedback2015,
  title = {Powerful {{Outflows}} and {{Feedback}} from {{Active Galactic Nuclei}}},
  author = {King, Andrew and Pounds, Ken},
  year = 2015,
  month = aug,
  journal = {ARA\&A},
  volume = {53},
  number = {1},
  pages = {115--154},
  issn = {0066-4146, 1545-4282},
  doi = {10.1146/annurev-astro-082214-122316},
  urldate = {2025-03-11},
}

@article{XRISM_Mission2025,
  title = {X-{{Ray Imaging}} and {{Spectroscopy Mission}}},
  author = {Tashiro, Makoto and Kelley, Richard and Watanabe, Shin and Maejima, Hironori and Reichenthal, Lillian and Toda, Kenichi and Hartz, Leslie and Santovincenzo, Andrea and Matsushita, Kyoko and Yamaguchi, Hiroya and Petre, Robert and Williams, Brian and Guainazzi, Matteo and Costantini, Elisa and Takei, Yoh and Ishisaki, Yoshitaka and Fujimoto, Ryuichi and {Henegar-Leon}, Joy and Sneiderman, Gary and Tomida, Hiroshi and Mori, Koji and Nakajima, Hiroshi and Terada, Yukikatsu and Holland, Matthew and Loewenstein, Michael and Miller, Eric and Sawada, Makoto and Kallman, Timothy and Kaastra, Jelle and Done, Chris and Enoto, Teruaki and Bamba, Aya and Corrales, Lia and Ueda, Yoshihiro and Kara, Erin and Zhuravleva, Irina and Fujita, Yutaka and Arai, Yoshitaka and Audard, Marc and Awaki, Hisamitsu and Ballhausen, Ralf and Baluta, Chris and Bando, Nobutaka and Behar, Ehud and Bialas, Thomas and {Boissay-Malaquin}, Rozenn and Brenneman, Laura and Brown, Gregory V and Chiao, Meng and Cumbee, Renata and {de~Vries}, Cor and {den~Herder}, Jan-Willem and D{\'i}az~Trigo, Mar{\'i}a and DiPirro, Michael and Dotani, Tadayasu and Carrero, Jacobo Ebrero and Ebisawa, Ken and Eckart, Megan and Eckert, Dominique and Eguchi, Satoshi and Ezoe, Yuichiro and Ferrigno, Carlo and Foster, Adam and Fukazawa, Yasushi and Fukushima, Kotaro and Furuzawa, Akihiro and Gallo, Luigi C and Garcia~Martinez, Javier and Gorter, Nathalie and Grim, Martin and Gu, Liyi and Hagino, Kouichi and Hamaguchi, Kenji and Hatsukade, Isamu and Hayashi, Katsuhiro and Hayashi, Takayuki and Hell, Natalie and {Hodges-Kluck}, Edmund and Horiuchi, Takafumi and Hornschemeier, Ann and Hoshino, Akio and Ichinohe, Yuto and Ikuta, Chisato and Iizuka, Ryo and Ishi, Daiki and Ishida, Manabu and Ishihama, Naoki and Ishikawa, Kumi and Ishimura, Kosei and Jaffe, Tess and Katsuda, Satoru and Kanemaru, Yoshiaki and Kenyon, Steven and Kilbourne, Caroline and Kimball, Mark and Kitamoto, Shunji and Kobayashi, Shogo and Kohmura, Takayoshi and Kubota, Aya and Leutenegger, Maurice A and Maeda, Yoshitomo and Markevitch, Maxim and Matsumoto, Hironori and Matsuzaki, Keiichi and McCammon, Dan and McLaughlin, Brian and McNamara, Brian and Mernier, Fran{\c c}ois and Miko, Joseph and Miller, Jon M and Minesugi, Kenji and Mitani, Shinji and Mitsuishi, Ikuyuki and Mizumoto, Misaki and Mizuno, Tsunefumi and Mukai, Koji and Murakami, Hiroshi and Mushotzky, Richard and Nakazawa, Kazuhiro and Natsukari, Chikara and Ness, Jan-Uwe and Nigo, Kenichiro and Nishiyama, Mari and Nobukawa, Kumiko and Nobukawa, Masayoshi and Noda, Hirofumi and Odaka, Hirokazu and Ogawa, Mina and Ogawa, Shoji and Ogorzalek, Anna and Okajima, Takashi and Okamoto, Atsushi and Ota, Naomi and Ozaki, Masanobu and Paltani, Stephane and Plucinsky, Paul and Porter, F Scott and Pottschmidt, Katja and Quero, Jose Antonio and Sasaki, Takahiro and Sato, Kosuke and Sato, Rie and Sato, Toshiki and Sato, Yoichi and Seta, Hiromi and Shida, Maki and Shidatsu, Megumi and Shigeto, Shuhei and Shipman, Russel and Shinozaki, Keisuke and Shirron, Peter and Simionescu, Aurora and Smith, Randall K and Soong, Yang and Suzuki, Hiromasa and Szymkowiak, Andrew and Takahashi, Hiromitsu and Takeo, Mai and Tamagawa, Toru and Tamura, Keisuke and Tanaka, Takaaki and Tanimoto, Atsushi and Terashima, Yuichi and Tsuboi, Yohko and Tsujimoto, Masahiro and Tsunemi, Hiroshi and Tsuru, Takeshi Go and Uchida, Hiroyuki and Uchida, Nagomi and Uchida, Yuusuke and Uchiyama, Hideki and Uno, Shinichiro and Vink, Jacco and Witthoeft, Michael and Wolfs, Rob and Yamada, Satoshi and Yamada, Shinya and Yamaoka, Kazutaka and Yamasaki, Noriko and Yamauchi, Makoto and Yamauchi, Shigeo and Yanagase, Keiichi and Yaqoob, Tahir and Yasuda, Susumu and Yoneyama, Tomokage and Yoshida, Tessei and Yukita, Mihoko},
  year = 2025,
  month = sep,
  journal = {PASJ},
  volume = {77},
  number = {Supplement\_1},
  pages = {S1-S9},
  issn = {0004-6264, 2053-051X},
  doi = {10.1093/pasj/psaf023},
  urldate = {2026-08-10},
  copyright = {https://creativecommons.org/licenses/by/4.0/},
}

@article{NuSTAR_Mission2013,
  title = {{{THE}} {{{\emph{NUCLEAR SPECTROSCOPIC TELESCOPE ARRAY}}}} ( {{{\emph{NuSTAR}}}} ) {{HIGH-ENERGY X-RAY MISSION}}},
  author = {Harrison, Fiona A. and Craig, William W. and Christensen, Finn E. and Hailey, Charles J. and Zhang, William W. and Boggs, Steven E. and Stern, Daniel and Cook, W. Rick and Forster, Karl and Giommi, Paolo and Grefenstette, Brian W. and Kim, Yunjin and Kitaguchi, Takao and Koglin, Jason E. and Madsen, Kristin K. and Mao, Peter H. and Miyasaka, Hiromasa and Mori, Kaya and Perri, Matteo and Pivovaroff, Michael J. and Puccetti, Simonetta and Rana, Vikram R. and Westergaard, Niels J. and Willis, Jason and Zoglauer, Andreas and An, Hongjun and Bachetti, Matteo and Barri{\`e}re, Nicolas M. and Bellm, Eric C. and Bhalerao, Varun and Brejnholt, Nicolai F. and Fuerst, Felix and Liebe, Carl C. and Markwardt, Craig B. and Nynka, Melania and Vogel, Julia K. and Walton, Dominic J. and Wik, Daniel R. and Alexander, David M. and Cominsky, Lynn R. and Hornschemeier, Ann E. and Hornstrup, Allan and Kaspi, Victoria M. and Madejski, Greg M. and Matt, Giorgio and Molendi, Silvano and Smith, David M. and Tomsick, John A. and Ajello, Marco and Ballantyne, David R. and Balokovi{\'c}, Mislav and Barret, Didier and Bauer, Franz E. and Blandford, Roger D. and Brandt, W. Niel and Brenneman, Laura W. and Chiang, James and Chakrabarty, Deepto and Chenevez, Jerome and Comastri, Andrea and Dufour, Francois and Elvis, Martin and Fabian, Andrew C. and Farrah, Duncan and Fryer, Chris L. and Gotthelf, Eric V. and Grindlay, Jonathan E. and Helfand, David J. and Krivonos, Roman and Meier, David L. and Miller, Jon M. and Natalucci, Lorenzo and Ogle, Patrick and Ofek, Eran O. and Ptak, Andrew and Reynolds, Stephen P. and Rigby, Jane R. and Tagliaferri, Gianpiero and Thorsett, Stephen E. and Treister, Ezequiel and Urry, C. Megan},
  year = 2013,
  month = may,
  journal = {ApJ},
  volume = {770},
  number = {2},
  pages = {103},
  issn = {0004-637X, 1538-4357},
  doi = {10.1088/0004-637X/770/2/103},
  urldate = {2026-08-10},
  copyright = {http://iopscience.iop.org/info/page/text-and-data-mining},
}

@article{XRISM_Xtend2025a,
  title = {Soft {{X-ray Imager}} of the {{Xtend}} System on Board {{XRISM}}},
  author = {Noda, Hirofumi and Mori, Koji and Tomida, Hiroshi and Nakajima, Hiroshi and Tanaka, Takaaki and Murakami, Hiroshi and Uchida, Hiroyuki and Suzuki, Hiromasa and Kobayashi, Shogo Benjamin and Yoneyama, Tomokage and Hagino, Kouichi and Nobukawa, Kumiko and Uchiyama, Hideki and Nobukawa, Masayoshi and Matsumoto, Hironori and Tsuru, Takeshi Go and Yamauchi, Makoto and Hatsukade, Isamu and Odaka, Hirokazu and Kohmura, Takayoshi and Yamaoka, Kazutaka and Yoshida, Tessei and Kanemaru, Yoshiaki and Hiraga, Junko and Dotani, Tadayasu and Ozaki, Masanobu and Tsunemi, Hiroshi and Sato, Jin and Takaki, Toshiyuki and Terada, Yuta and Miyazaki, Keitaro and Kusunoki, Kohei and Otsuka, Yoshinori and Yokosu, Haruhiko and Yonemaru, Wakana and Ichikawa, Kazuhiro and Nakano, Hanako and Takemoto, Reo and Matsushima, Tsukasa and Urase, Reika and Kurashima, Jun and Fuchi, Kotomi and Hayakawa, Kaito and Fukuda, Masahiro and Kamei, Takamitsu and Asahina, Yoh and Inoue, Shun and Amano, Yuki and Aoki, Yuma and Ito, Yamato and Kamatani, Tomoya and Takayama, Kouta and Sako, Takashi and Yoshimoto, Marina and Shima, Kohei and Higuchi, Mayu and Ninoyu, Kaito and Aoki, Daiki and Tsunomachi, Shun and Hayashida, Kiyoshi},
  year = 2025,
  month = sep,
  journal = {PASJ},
  volume = {77},
  number = {Supplement\_1},
  pages = {S10-S22},
  issn = {0004-6264, 2053-051X},
  doi = {10.1093/pasj/psaf011},
  urldate = {2026-08-10},
  copyright = {https://creativecommons.org/licenses/by/4.0/},
}

@article{XRISM_Resolve2025,
  title = {Resolve Instrument Onboard the {{X-Ray Imaging}} and {{Spectroscopy Mission}}},
  author = {Kelley, Richard L. and Ishisaki, Yoshitaka and Costantini, Elisa and Awaki, Hisamitsu and Balleza, Jesus C. and Barnstable, Kim R. and Bialas, Thomas G. and {Boissay-Malaquin}, Rozenn and Brown, Gregory V. and Canavan, Edgar R. and Timothy M, Carnahan and Chiao, Meng P. and Comber, Brian J. and Cumbee, Renata S. and den Herder, Jan-Willem and Dercksen, Johannes and de Vries, Cor P. and DiPirro, Michael J. and Eckart, Megan E. and Ezoe, Yuichiro and Ferrigno, Carlo and Fujimoto, Ryuichi and Gorter, Nathalie and Graham, Steven M. and Grim, Martin and Hartz, Leslie S. and Hayakawa, Ryota and Hayashi, Takayuki and Hell, Natalie and Ichinohe, Yuto and Ishi, Daiki and Ishida, Manabu and Ishikawa, Kumi and James, Bryan L. and Kanemaru, Yoshiaki and Kenyon, Steven J. and Kilbourne, Caroline A. and Kimball, Mark and Kitamoto, Shunji and Leutenegger, Maurice A. and Maeda, Yoshitomo and McCammon, Dan and McLaughlin, Brian J. and Miko, Joseph J. and van der Meer, Erik and Mizumoto, Misaki and Noda, Hirofumi and Okajima, Takashi and Okamoto, Atsushi and Paltani, Stephane and Porter, Frederick S. and Reichenthal, Lillian S. and Sato, Kosuke and Sato, Toshiki and Sato, Yoichi and Sawada, Makoto and Shinozaki, Keisuke and Shipman, Russell and Shirron, Peter J. and Sneiderman, Gary A. and Soong, Yang and Szymkiewicz, Richard and Szymkowiak, Andrew E. and Takei, Yoh and Takeo, Mai and Tamura, Keisuke and Tsujimoto, Masahiro and Uchida, Yuusuke and Wasserzug, Stephen and Witthoeft, Michael C. and Wolfs, Rob and Yamada, Shinya and Yamasaki, Noriko Y. and Yasuda, Susumu},
  year = 2025,
  month = nov,
  journal = {JATIS},
  volume = {11},
  number = {4},
  pages = {042026},
  publisher = {SPIE},
  issn = {2329-4124},
  doi = {10.1117/1.JATIS.11.4.042026},
  urldate = {2026-08-10},
}

@article{XMMNewton2001,
  title = {{{XMM-Newton}} Observatory - {{I}}. {{The}} Spacecraft and Operations},
  author = {Jansen, F. and Lumb, D. and Altieri, B. and Clavel, J. and Ehle, M. and Erd, C. and Gabriel, C. and Guainazzi, M. and Gondoin, P. and Much, R. and Munoz, R. and Santos, M. and Schartel, N. and Texier, D. and Vacanti, G.},
  year = 2001,
  month = jan,
  journal = {A\&A},
  volume = {365},
  number = {1},
  pages = {L1-L6},
  publisher = {EDP Sciences},
  issn = {0004-6361, 1432-0746},
  doi = {10.1051/0004-6361:20000036},
  urldate = {2026-08-10},
  copyright = {\copyright{} ESO, 2001},
}

\appendix 

\section{The continuum variability of the X-ray source}\label{appendix:continuum_variability}
\begin{figure}[tbp]
  \begin{center}
    \includegraphics[width=7cm]{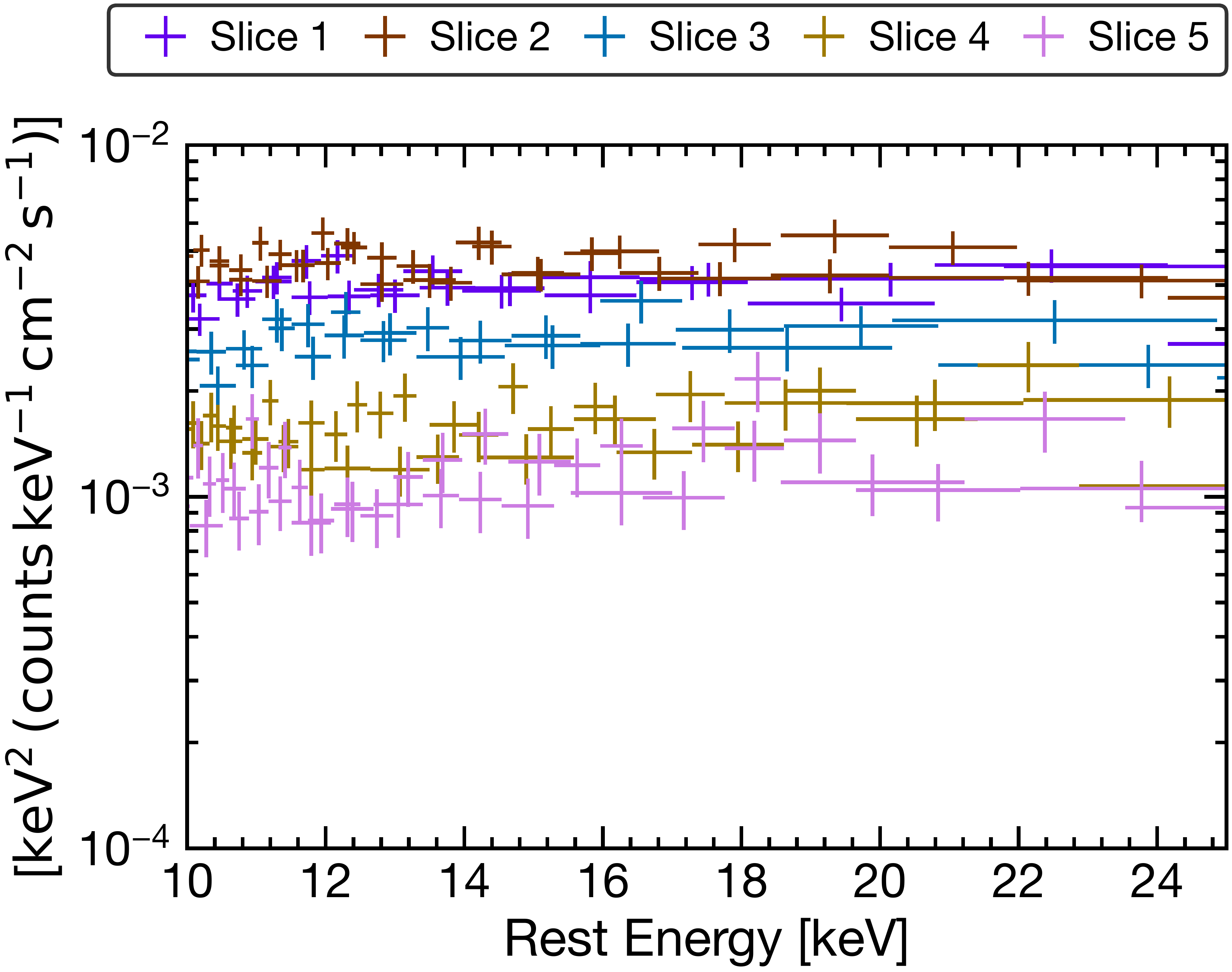}
  \end{center}
  \caption{\nustar\ 10--25~keV spectra for time slices 1--5.
  {Alt text: Line graph of five NuSTAR spectra. Rest energy spans 10 to 25 kilo electron volt on the x axis, and the intensity axis is logarithmic from about ten to the minus four to ten to the minus two. The five time slices overlap within errors and show no clear spectral shape difference. }
  }\label{fig:5period_nustar_eed}
\end{figure}

Figure~\ref{fig:5period_nustar_eed} shows the \nustar\ 10--25~keV spectra for slices 1--5 defined in Fig.~\ref{fig:7period_lightcurve}. The spectra exhibit no significant variability in this energy band, which is dominated by the power-law continuum. This justifies our decision to link the photon index of the power-law component across all time slices in the time-sliced spectral fitting.
\revistwomaj{
\section{Rationale for the time-sliced spectral-fitting strategy}\label{appendix:time_sliced_fitting_strategy}
In the time-sliced spectral analysis, we linked the ionization parameter $\xi$ and outflow velocity $v_\mathrm{out}$ of the low-ionization UFO across all seven slices. To assess this assumption, we performed a test fit in which both parameters were allowed to vary independently among the slices. Figures~\ref{fig:7period_velocity_var} and \ref{fig:7period_rlogxi_var} show the resulting best-fit values of $v_\mathrm{out}$ and $\log \xi$, respectively. Apart from slice 5 for $v_\mathrm{out}$ and slice 6 for $\log \xi$, the values are broadly consistent with being constant within their statistical uncertainties. Because an isolated deviation in a single slice does not provide compelling evidence for physical variability, we adopted the linked-parameter configuration for the main analysis.
\revisthree{Because the possible variation in $\log \xi$ shown in Fig.~\ref{fig:7period_rlogxi_var} may raise concern that the inferred $C_\mathrm{f}$ variability is caused by degeneracy between the two parameters, we calculated their 90\% confidence contours with the \textsc{steppar} command in \textsc{xspec}, using the test fit in which $\log \xi$ was allowed to vary independently among the slices (Fig.~\ref{fig:covfrac_rlogxi_contours}). The contours show no clear common diagonal orientation that would indicate a strong and systematic degeneracy between $\log \xi$ and $C_\mathrm{f}$. Although the contours overlap for several slices, the covering fractions broadly follow the same temporal trend as that obtained in the main fit (Fig.~\ref{fig:7period_covfrac_var}), and this evolution is not aligned with a common $\log \xi$--$C_\mathrm{f}$ covariance direction. These results do not support the interpretation that the apparent covering-fraction changes are produced solely by a $\log \xi$--$C_\mathrm{f}$ degeneracy.}
\begin{figure}[tbp]
  \begin{center}
  \includegraphics[width=6cm]{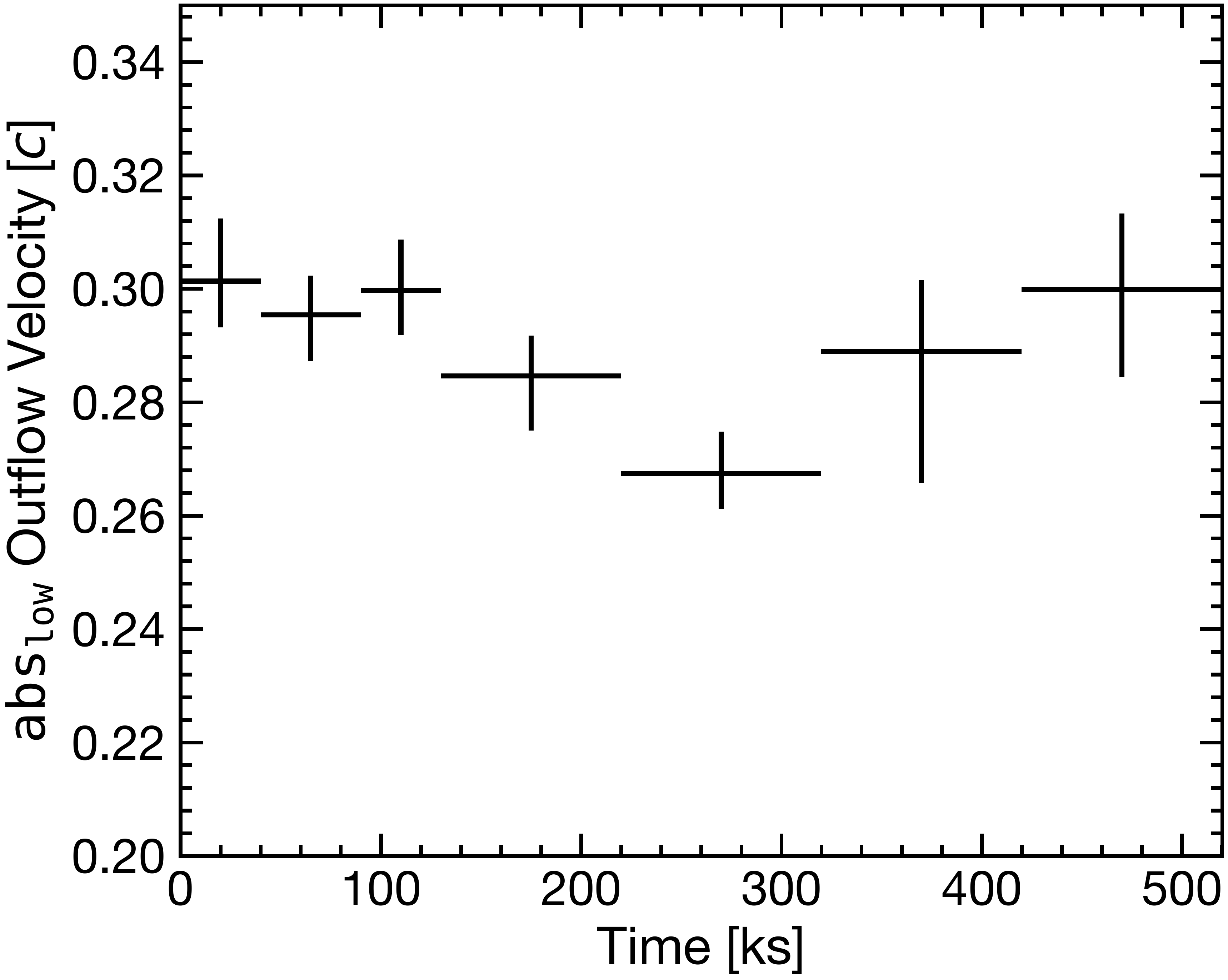}
  \end{center}
  \caption{
    \revistwomaj{Outflow velocity of the low-ionization UFO obtained from the test fit in which $v_\mathrm{out}$ was allowed to vary independently among the seven time slices. The horizontal and vertical error bars represent the duration of each time slice and the statistical uncertainty, respectively.
    {Alt text: Graph of low-ionization UFO outflow velocity for seven time slices from zero to 520 kiloseconds. Apart from slice 5, the points are nearly constant around 0.30 times the speed of light.}}}
  \label{fig:7period_velocity_var}
\end{figure}
\begin{figure}[tbp]
  \begin{center}
  \includegraphics[width=6cm]{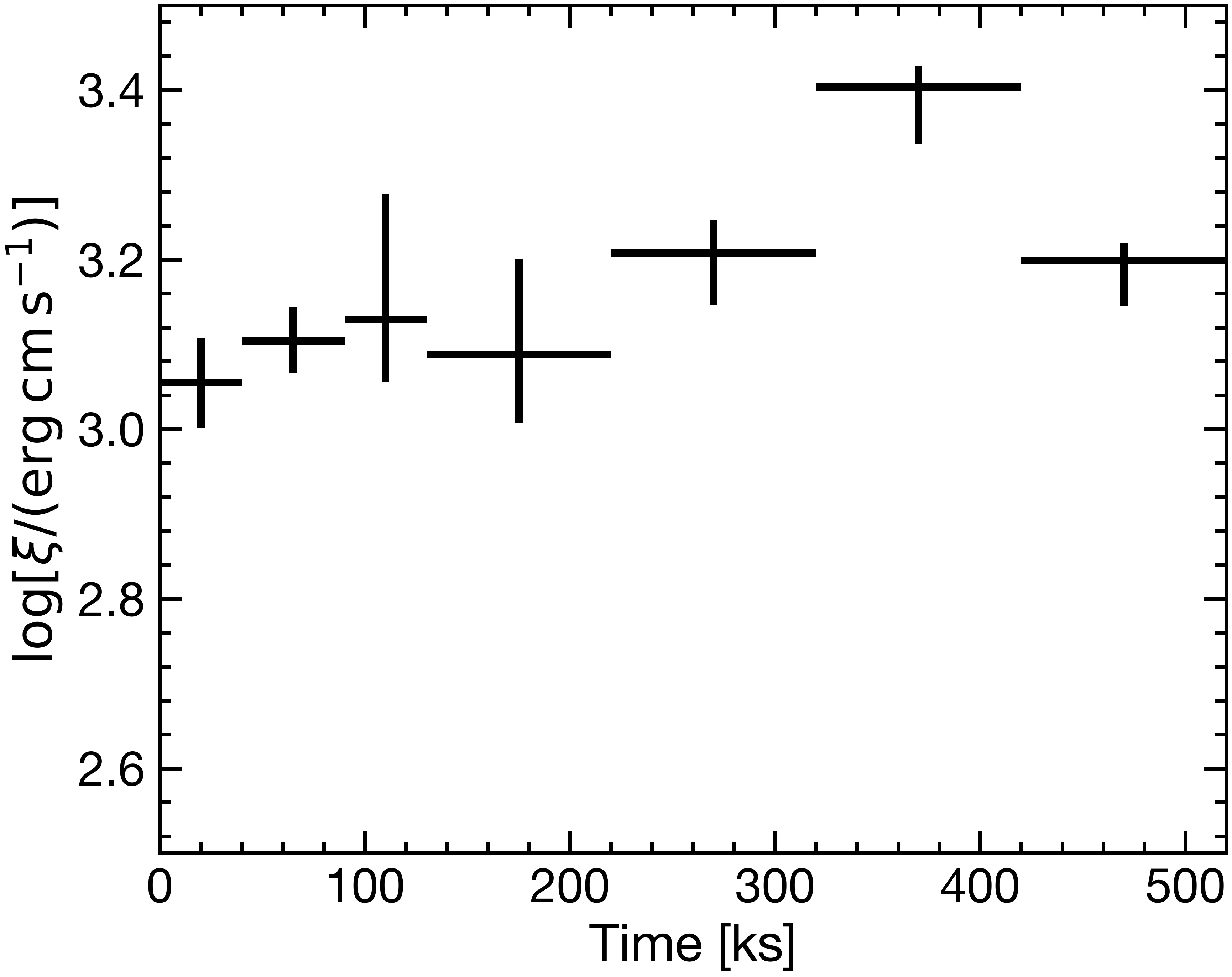}
  \end{center}
  \caption{ 
    \revistwomaj{
    Ionization parameter of the low-ionization UFO obtained from the test fit in which $\log \xi$ was allowed to vary independently among the seven time slices. The horizontal and vertical error bars represent the duration of each time slice and the statistical uncertainty, respectively.
    {Alt text: Graph of the low-ionization UFO ionization parameter for seven time slices from zero to 520 kiloseconds. Apart from slice 6, the points are nearly constant around $\log \xi \simeq 3.1$--$3.2$.}}}
  \label{fig:7period_rlogxi_var}
\end{figure}
\revisthree{
\begin{figure}[tbp]
  \begin{center}
  \includegraphics[width=8cm]{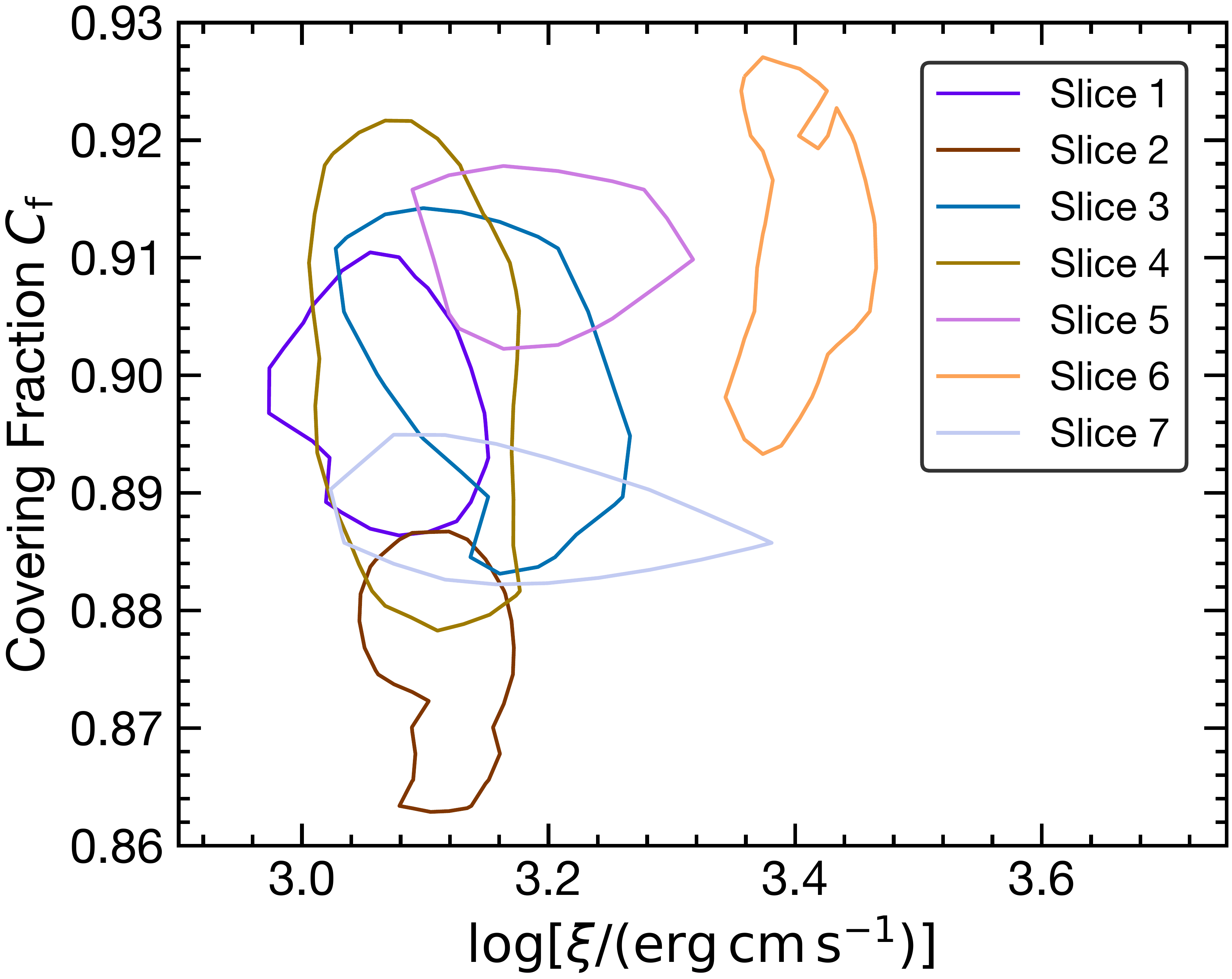}
  \end{center}
  \caption{
  \revisthree{90\% confidence contours between the ionization parameter $\log \xi$ and covering fraction $C_\mathrm{f}$ of the low-ionization UFO for the seven time slices, obtained from the test fit in which $\log \xi$ was allowed to vary independently among the slices.
  {Alt text: Confidence-contour plot of covering fraction versus ionization parameter for seven time slices. Several contours overlap around log xi of 3.0 to 3.2 and covering fractions of 0.88 to 0.92, whereas the slice 6 contour is separated at a higher log xi of about 3.4 and a high covering fraction.}}}
  \label{fig:covfrac_rlogxi_contours}
\end{figure}
}
}
\revistwomaj{
\section{The variability of the high-ionization UFO absorption}\label{appendix:high_ionization_UFO_variability}
Table~\ref{tab:7term_fitting} and Fig.~\ref{fig:7period_column_var} show possible variability in the total column density of the high-ionization UFO across the seven time slices. The total column density generally increases through slice 6 and then decreases in slice 7. We note that $C_\mathrm{f}$ applies to both $\mathtt{abs}_{\tt low}$ and $\mathtt{abs}_{\tt high}$ in equation~(\ref{eq:model1}). Consequently, the total column density of $\mathtt{abs}_{\tt high}$ is coupled with $C_\mathrm{f}$, and the variation in $C_\mathrm{f}$ may affect the inferred column-density trend. However, $C_\mathrm{f}$ varies by at most $\sim0.05$, or about 5 percentage points, whereas the total column density changes by a factor of approximately three, from $\sim\num{2e23}$ to $\sim\num{6e23}\,\mathrm{cm^{-2}}$. The larger fractional variation in the total column density suggests that its trend is unlikely to be caused solely by its coupling with $C_\mathrm{f}$. \par
To examine the variability of the high-ionization UFO absorption lines with minimal dependence on the spectral model, we also calculated a data-driven equivalent width using only the \xtend\ data. We defined the equivalent width as
\begin{equation}\label{eq:high_ufo_equivalent_width}
  \mathrm{EW} = \sum_i \Delta E_i \left(1-\frac{D_i}{C_i}\right),
\end{equation}
where $i$ denotes the energy-bin index, $\Delta E_i$ is the width of bin $i$, $D_i$ is the rest-frame \xtend\ count-rate density corrected for the effective area, in units of $\mathrm{counts\,s^{-1}\,cm^{-2}}$, and $C_i$ is the corresponding continuum count-rate density predicted from a detector-response-convolved $\mathtt{tbabs}\times\mathtt{pow}$ model. The continuum-model parameters were fixed at their best-fit values from the main analysis. The equivalent-width uncertainties were calculated by propagating only the Poisson uncertainties of the data and were converted to 90\% confidence intervals. We did not include uncertainties in the continuum-model parameters. We evaluated equation~(\ref{eq:high_ufo_equivalent_width}) over the 8--10~keV rest-frame band, which contains the Fe-K absorption features. The resulting equivalent widths for the seven time slices are shown in Fig.~\ref{fig:7period_eqwidth_var}. The equivalent width shows a similar trend to the total column density, with an increase through slice 7. This consistency between the model-dependent column density and the model-independent equivalent width supports the reality of the observed variability in the high-ionization UFO absorption.\par
\revisthree{Possible explanations for the observed strengthening of the high-ionization UFO absorption include (1) variations in the intrinsic column density or covering fraction of the high-ionization clumps and (2) changes in the ionization state of the high-ionization UFO. The overall increase in opacity from slice 2 to slice 7, as the continuum flux declines following the flare, is qualitatively consistent with a decrease in the ionization state in response to the declining ionizing flux, which would strengthen the Fe-K absorption lines. Although the column density and equivalent width in slices 1 and 2 are consistent despite the higher continuum flux in slice 2, this may be attributable to their statistical uncertainties. Variations in the intrinsic column density or covering fraction are also plausible, and the present data do not allow us to distinguish among these mechanisms.}\par
We note that the inferred variability in $C_\mathrm{f}$ is not driven by the Fe-K structure. We confirmed this by fixing $\mathtt{abs}_{\tt high}$ and $\mathtt{emiss}_{\tt high}$ to their time-averaged best-fit values and repeating the time-sliced fitting after excluding the Fe-K band($7\text{--}10$~keV in the observed frame). The resulting covering fractions are consistent with those from the main analysis and show the same temporal trend.
\begin{figure}[tbp]
  \begin{center}
  \includegraphics[width=6cm]{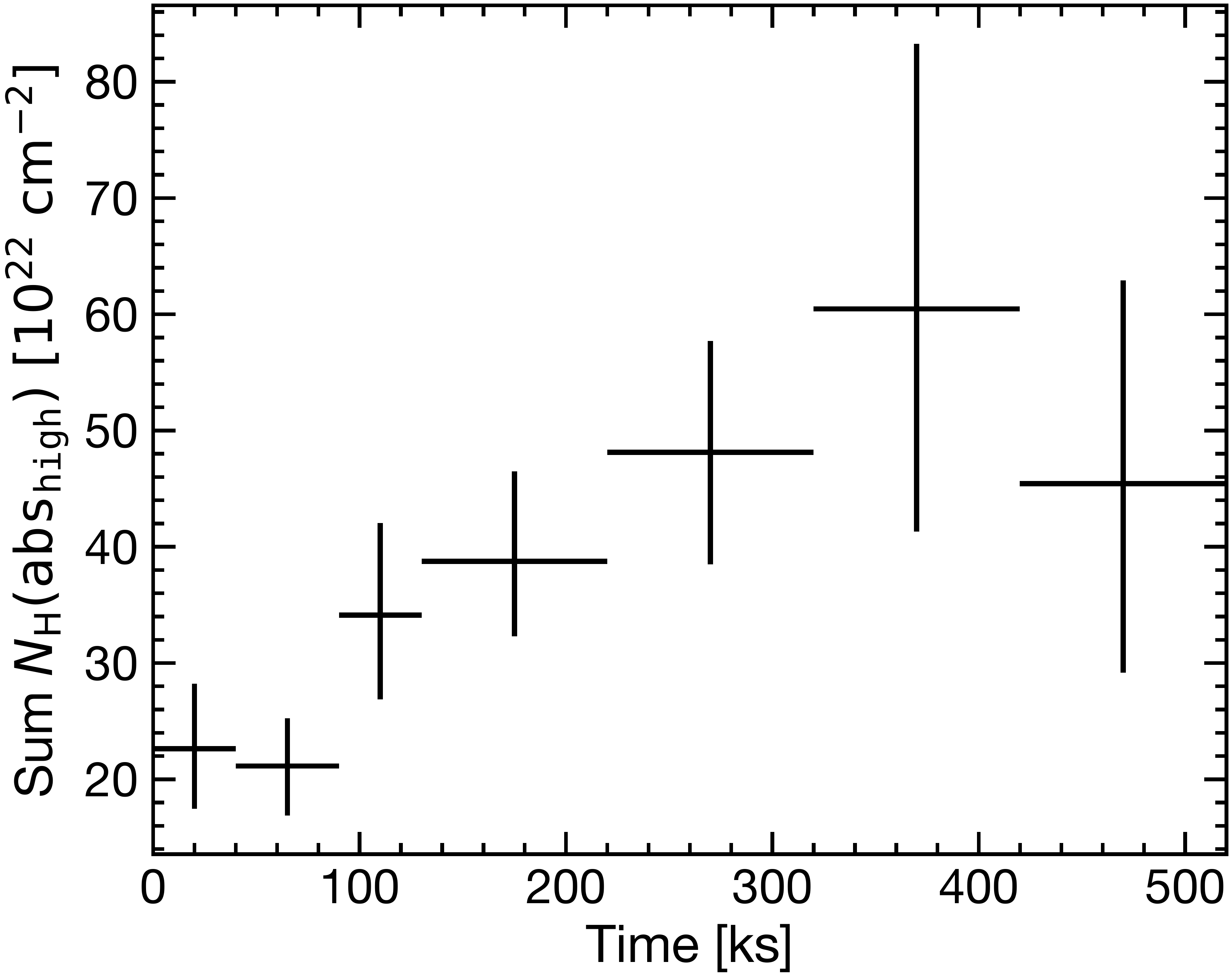}
  \end{center}
  \caption{ 
    \revistwomaj{
    Variation in the total column density of the five high-ionization UFO absorption components across the seven time slices. The horizontal and vertical error bars represent the duration of each time slice and the statistical uncertainty, respectively.
    {Alt text: Graph with time from zero to 520 kiloseconds on the x axis and total column density from about 15 to 85 times ten to the twenty-two per square centimeter on the y axis. The values generally rise through slice 6 and decrease in slice 7.}}}
  \label{fig:7period_column_var}
\end{figure} 
\begin{figure}[tbp]
  \begin{center}
  \includegraphics[width=6cm]{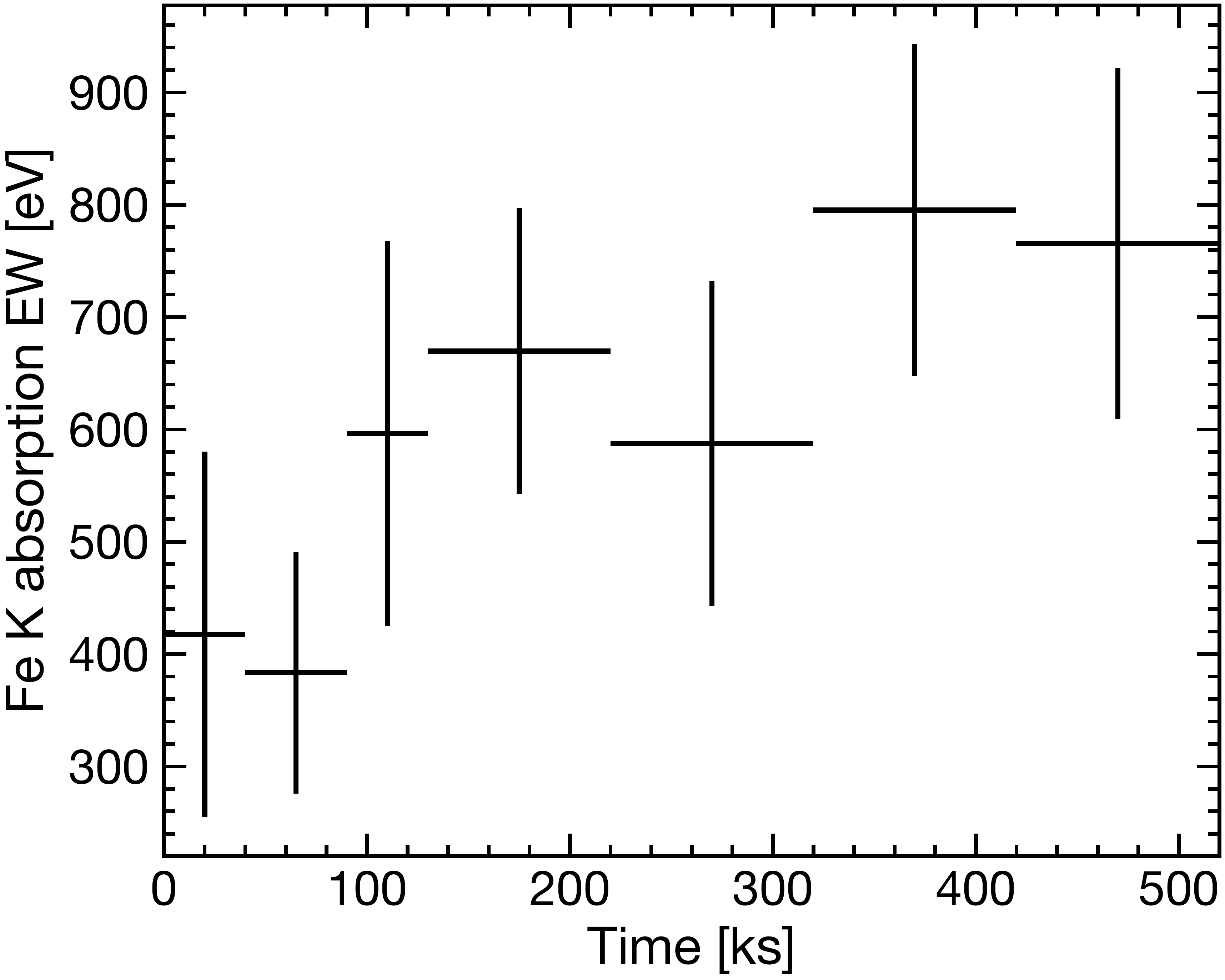}
  \end{center}
  \caption{ 
    \revistwomaj{Variation in the equivalent width of the five high-ionization UFO absorption components across the seven time slices. The horizontal and vertical error bars represent the duration of each time slice and the statistical uncertainty, respectively.
    {Alt text: Graph with time from zero to 520 kiloseconds on the x axis and equivalent width from about 15 to 85 times ten to the twenty-two per square centimeter on the y axis. The values generally rise through slice 7.}}}
  \label{fig:7period_eqwidth_var}
\end{figure} 
}
\revistwomaj{
\section{The variability of the low-ionization emission component}\label{appendix:low_ionization_emission}
Table~\ref{tab:7term_fitting} and Fig.~\ref{fig:7period_emiss_low_norm_var} show that the normalization of the low-ionization emission component, $\mathtt{emiss}_{\tt low}$, varies during the observation. In particular, the fitted normalization increases from $<\num{0.3e-4}$ in slice 2 to $\num{2.1(5)e-4}$ in slice 3. If this variability is real, the normalization rises nominally by a factor of $\gtrsim 7$ within $\lesssim 90$~ks. This timescale constrains the light-crossing size of the emitting region to $D_\mathrm{emiss}\lesssim 40\,R_\mathrm{g}$ for $M_\mathrm{BH}\sim\num{5e8}\,M_\odot$ \citep{GravityCollab_2023}. If, in addition, the increase in slice 3 is a response to the X-ray flare in slice 2 (Fig.~\ref{fig:7period_lightcurve}), the separation between the X-ray corona and the emitting region is also constrained to $\lesssim 40\,R_\mathrm{g}$. This favors an additional compact component close to the X-ray corona, rather than interpreting all of $\mathtt{emiss}_{\tt low}$ as emission from the low-ionization UFO located at $r \gtrsim \num{4e3}\,R_\mathrm{g}$. One possible origin of this additional component is the soft excess, whose physical origin remains uncertain. A relevant possibility for a nearly face-on super- or near-Eddington AGN is warm Comptonization: disk photons may be Comptonized in warm, optically thick material associated with a puffed-up inner-disk surface \citep{Jin_softexcess_2017}, a warm corona \citep{Done_softexcess_2012,Petrucci_softexcess_2018}, or an optically thick wind photosphere \citep{KingPounds_UFO2003,Pounds2003}. Such a region could be both compact and close to the X-ray corona, consistent with the constraints above. Determining the physical origin of $\mathtt{emiss}_{\tt low}$ requires more detailed modeling and is beyond the scope of this paper; we leave this question for future work.\par
\begin{figure}[tbp]
  \begin{center}
    \includegraphics[width=6cm]{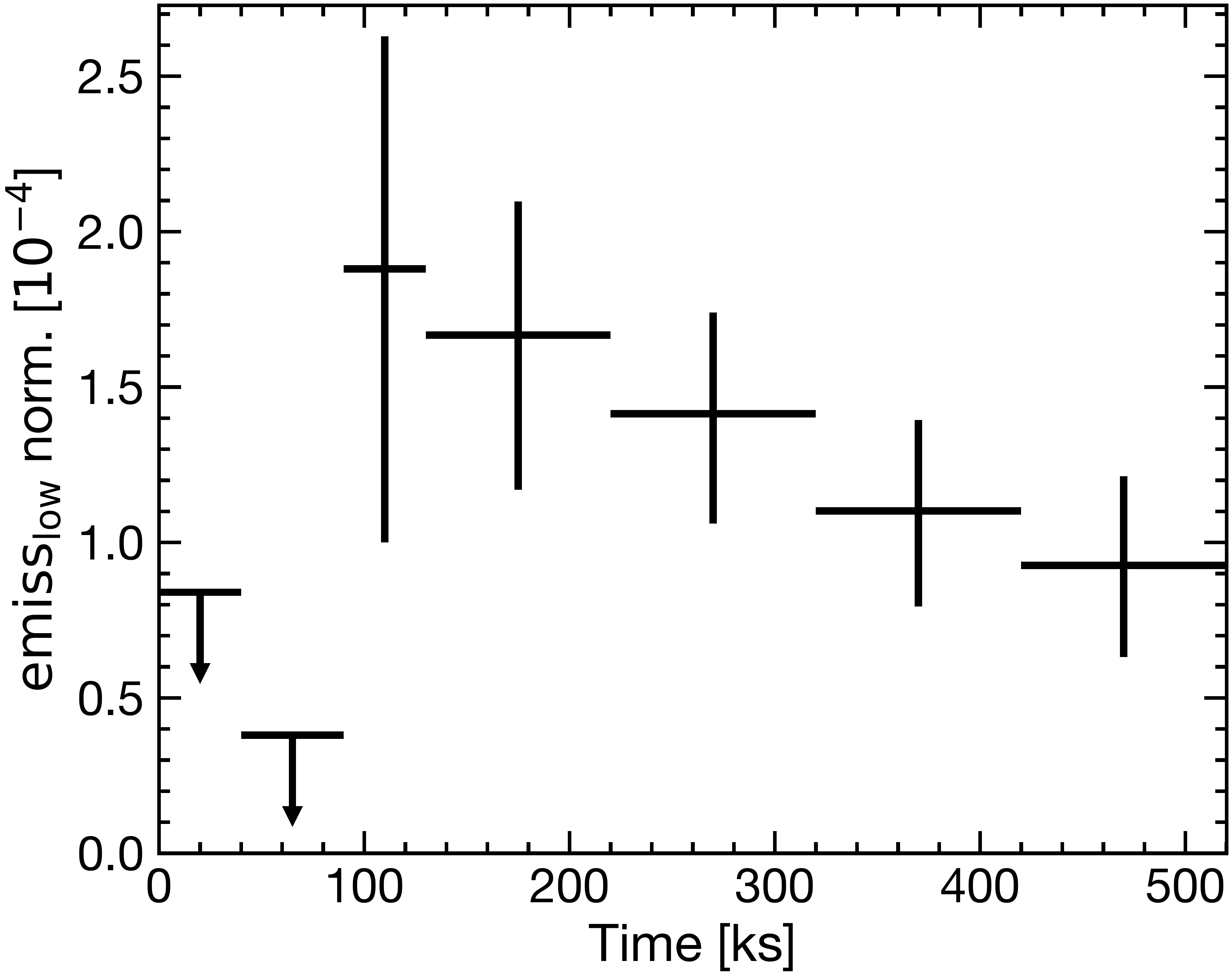}
  \end{center}
  \caption{\revistwomaj{Variation in the normalization of the low-ionization emission component across the seven time slices. The horizontal error bars represent the duration of each time slice, and the vertical error bars represent the $1\sigma$ confidence intervals.
  {Alt text: Graph of the low-ionization emission normalization in seven time slices from zero to 520 kiloseconds. The normalization is low in slices 1 and 2, rises in slice 3, and then gradually decreases through slice 7. }}
  }\label{fig:7period_emiss_low_norm_var}
\end{figure}
\begin{figure}[tbp]
  \begin{center}
    \includegraphics[width=6cm]{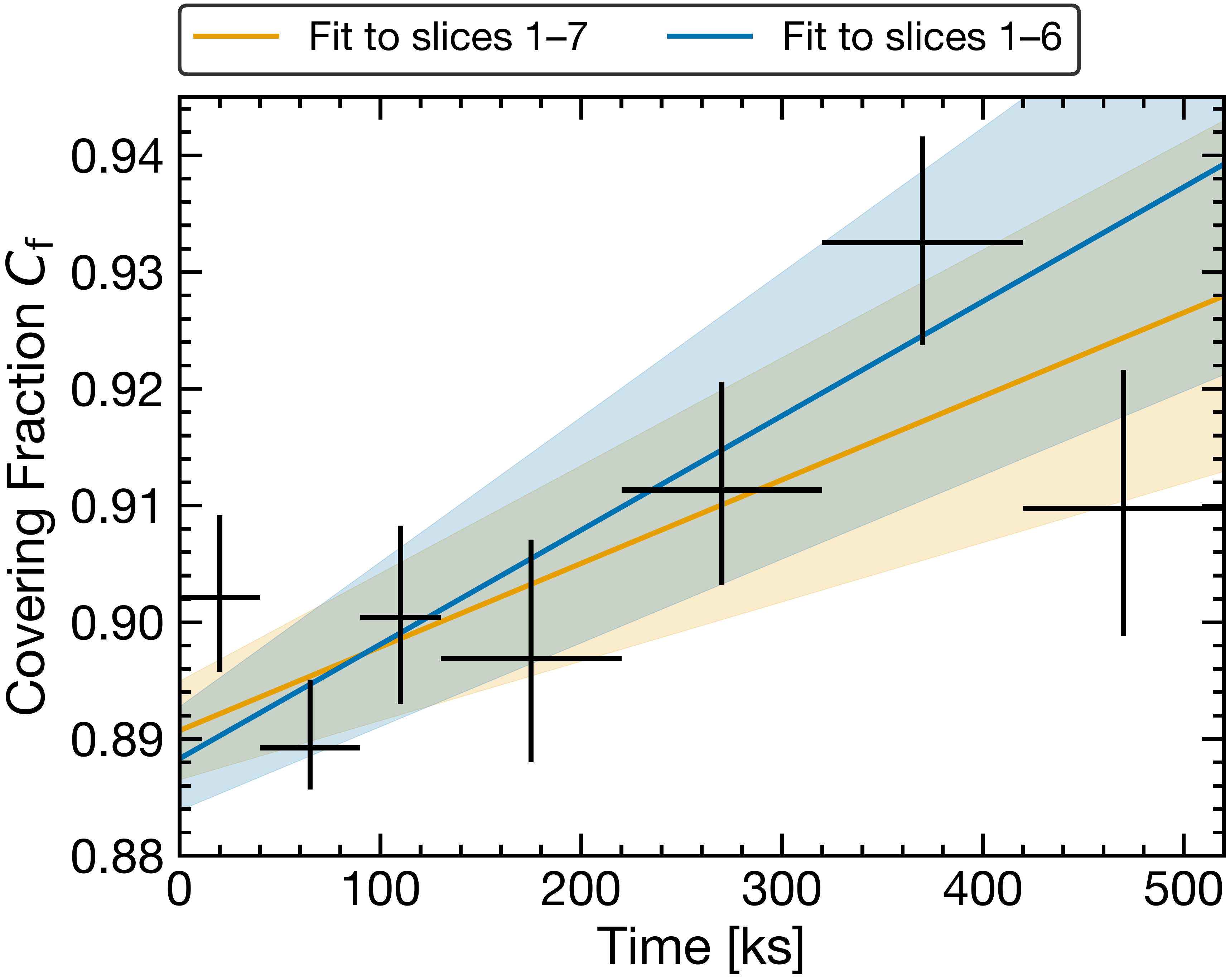}
  \end{center}
  \caption{\revistwomaj{Same as Fig.~\ref{fig:7period_covfrac_var}, but with the normalization of the low-ionization emission component fixed across all time slices.
  {Alt text: Graph of the low-ionization UFO covering fraction in seven time slices from zero to 520 kiloseconds, obtained with the low-ionization emission normalization fixed across all slices. The covering fraction generally increases through slice 6 and decreases in slice 7. A best-fit increasing linear trend and its uncertainty range are shown. }}
  }\label{fig:7period_emiss_low_norm_var_emisslowfix}
\end{figure}
We also tested whether the inferred covering-fraction variability is affected by allowing the normalization of $\mathtt{emiss}_{\tt low}$ to vary. Fixing this normalization to be the same across all time slices worsens the fit by $\Delta C = 37.1$ for $\Delta\mathrm{d.o.f.}=6$, corresponding to a formal significance of $\sim 4.8\sigma$ for variability in the emission component according to Wilks' theorem. In this fit, the common normalization is $\kappa=\num{1.09(30:29)e-4}$, which corresponds to a large global wind covering factor, $f_\mathrm{cov}\equiv\Omega/2\pi\sim1$ (i.e., a near-side solid angle of $\Omega\sim2\pi$), using equation~(4) of \citet{Firstpaper2025}. Figure~\ref{fig:7period_emiss_low_norm_var_emisslowfix} shows the covering fractions obtained from the fit with the normalization fixed. The increasing trend in the covering fraction persists, with a slope of $\qty{6.8(11)e-5}{\per\kilo\s}$ when all slices are included. The covering-fraction variability is therefore unlikely to be an artifact of the variable $\mathtt{emiss}_{\tt low}$ normalization, and the distance constraint on the low-ionization UFO derived from this variability remains valid.
}
\end{document}